\documentclass[useAMS,usenatbib]{mnras}
\usepackage{color}
\usepackage{graphicx}
\usepackage{lscape}
\usepackage{amssymb,amsmath}
\usepackage{pifont}
\usepackage{url}
\usepackage{float}

\title[Light curve analysis of Cepheid Variable]
{A Comparative Study of Multiwavelength Theoretical and Observed Light Curves of Cepheid Variables}
\author[Bhardwaj et al.]{Anupam Bhardwaj$^{1,2}$\thanks{E-mail:
anupam.bhardwajj@gmail.com; abhardwaj@eso.org}, Shashi M. Kanbur$^3$, Marcella Marconi$^4$, Marina Rejkuba$^2$, 
\newauthor 
Harinder P. Singh$^1$ and Chow-Choong Ngeow$^5$\\
\vspace{1pt}\\
1. Department of Physics \& Astrophysics, University of Delhi, Delhi 110007, India \\
2. European Southern Observatory, Karl-Schwarzschild-Stra\ss e 2, 85748, Garching, Germany\\
3. State University of New York, Oswego, NY 13126, USA\\
4. INAF-Osservatorio astronomico di Capodimonte, Via Moiariello 16, 80131 Napoli, Italy\\
5. Graduate Institute of Astronomy, National Central University, Jhongli 32001, Taiwan \\
}
\begin{document}

\date{Accepted 2016 December 10. Received 2016 December 10; in original form 2016 October 19}

\pagerange{\pageref{firstpage}--\pageref{lastpage}} \pubyear{2016}

\maketitle

\label{firstpage}

\begin{abstract}
We analyse the theoretical light curves of Cepheid variables at optical ({\it UBVRI}) and near-infrared ({\it JKL}) wavelengths using the Fourier decomposition and principal component analysis methods. The Cepheid light curves are based on the full amplitude, nonlinear, convective hydrodynamical models for chemical compositions representative of Cepheids in the Galaxy ($Y=0.28$, $Z=0.02$), Large Magellanic Cloud ($Y=0.25$, $Z=0.008$) and Small Magellanic Cloud ($Y=0.25$, $Z=0.004$). We discuss the variation of light curve parameters with different compositions and mass-luminosity levels as a function of period and wavelength, and compare our results with observations. For a fixed composition, the theoretical amplitude parameters decrease while the phase parameters increase with wavelength, similar to the observed Fourier parameters. The optical amplitude parameters obtained using canonical mass-luminosity Cepheid models, exhibit a large offset with respect to the observations for periods between 7-11 days, when compared to the non-canonical mass-luminosity levels. 
The central minimum of the Hertzsprung progression for amplitude parameters, shifts to the longer periods with decrease/increase in metallicity/wavelength for both theoretical and observed light curves. The principal components for Magellanic Clouds Cepheid models are consistent with observations at optical wavelengths. We also observe two distinct populations in the first principal component for optical and near-infrared wavelengths while $J$-band contributes to both populations. Finally, we take into account the variation in the convective efficiency by increasing the adopted mixing length parameter from the standard 1.5 to 1.8. This results in a zero-point offset in the bolometric mean magnitudes and in amplitude parameters (except close to 10 days), reducing the systematically large difference in theoretical amplitudes.
 \end{abstract}

\begin{keywords}
stars: variables: Cepheids, stars: evolution, (galaxies:) Magellanic Clouds.
\end{keywords}

\section{Introduction}
\label{sec:intro}

Classical Cepheids and RR Lyrae variables provide a unique opportunity for a rigorous comparison of pulsation properties from the theoretical models with observations, leading to important constraints for the theory of stellar pulsation and evolution \citep{cox1980a}. Cepheid variables are extensively used as standard candles in extra-galactic distance determination through Period-Luminosity \citep[P-L,][]{leavitt} and Period-Luminosity-Color (P-L-C) relations. They play a vital role in cosmic distance ladder for the determination of Hubble constant to a percent-level precision \citep{riess16}. 

The light curves of Cepheid variables with period from about 6 to $\sim16$ days days show a bump \citep[Hertzsprung progression, hereafter HP,][]{hertzsprung1926}, which varies with period. This secondary feature appears on the descending branch in short period Cepheids, reaches the maximum light around a period of 10 days and shifts to the earlier phases for periods longer than 10 days. The first quantitative study of the light curve structure of Cepheid variables was carried out by \citet{slee81} using the Fourier analysis method. \citet{simon1983} carried out a comparison of the observed light and velocity curves of Classical Cepheids with theoretical light curves and suggested that the phase lag obtained from Fourier decomposition is most useful for comparison with observations.
\vspace{7pt}

Theoretical investigations of pulsation properties of Classical Cepheids were carried out in a series of papers \citep[][and references therein]{bono1998, bono1999b, bono2000d, caputo2000b, caputo2000c}. \citet{bono1999b} used the mass and luminosities provided by stellar evolutionary calculations as input parameters to pulsation models together with different chemical compositions to generate theoretical light curves of Cepheid variables and studied theoretical P-L and P-L-C relations. Later, \citet{caputo2000b} derived these relations over multiple wavelengths and found a fair agreement with observed Galactic and Magellanic Clouds (MC) Cepheids. The Cepheid pulsation models and the effects of chemical abundances on the P-L and Period-Wesenheit (P-W) relations has been a subject of many studies in the past \citep[see,][and references therein]{fiorentino2002, marconi2005, fiorentino2007, marconi2013}. For example, \citet{bono2010} found that the slope of the P-L relations become steeper and less dependent on metallicity, from optical to near-infrared (NIR) bands. These studies of the P-L and P-W relations follow classical approach of using mean light properties of Cepheid variables which, being the average over pulsation phase, neglects various physical features of the light curve structure.
\vspace{7pt}

In order to analyse the Cepheids light curves, \citet{bono2000d} carried out an extensive study on the HP for Classical Cepheids and found the central period of the progression around 11 days for models with Z=0.008 and Y=0.25, in a good agreement with the empirical value derived using the Fourier parameters of Large Magellanic Cloud (LMC) Cepheid light curves. \citet{bono2002} also used pulsation models to investigate the bump in the $I$-band light curves of two Cepheids in the LMC. However, so far no rigorous comparison of light curve parameters at multiple wavelengths for different compositions was carried out. 
\vspace{7pt}

Recently, we quantitatively analysed the observed Cepheid light curves in the Galaxy and LMC at multiple wavelengths using the Fourier decomposition technique \citep{bhardwaj2015a}. We found various features in the amplitude, phase, skewness and acuteness parameters as a function of period and wavelength. For example, the mean Fourier amplitude parameter values are distinctly separated around 20 days for wavelengths longer than $J$ vs. optical bands. Furthermore, the central period of the HP was found to vary as a function of wavelength and metallicity. In this work, we present an investigation of theoretical Cepheid light curves generated using Cepheid pulsation models and compare them with observed light curve parameters from \citet{bhardwaj2015a}. The aim of this work is to form a basis to explore strong constraints for stellar pulsation models by an extensive comparison of theoretical and observed light curves.
\vspace{7pt}

The paper is structured as follows. We discuss the theoretical framework used to generate the light curves for Cepheids at multiple wavelengths in Section~\ref{sec:theory}. In Section~\ref{sec:fourier}, we discuss the application of the Fourier decomposition on the theoretical light curves and the variation of light curve parameters with period, wavelength and metallicity. We also compare the theoretical and observed Fourier parameters. In Section~\ref{sec:pca}, we discuss the principal component analysis
of the theoretical light curves together with the observed light curves. We summarize our findings from this study in Section~\ref{sec:discuss}.

\section{Theoretical and Observational Framework}
\label{sec:theory}

We use full amplitude, nonlinear, convective Cepheid models to generate the theoretical light curves. The computed models adopt the same hydrodynamical code and physical and numerical assumptions as in \citet[][and references therein]{marconi2013}. These models are able to follow the limit cycle stability of both fundamental (FU) and first-overtone (FO) models and include a non-local, time-dependent treatment of convection to properly follow the coupling between dynamical and convective velocities along a pulsation cycle. For each fixed chemical composition and mass we consider the luminosity predicted by the canonical mass-luminosity (ML) relation by \citet{bonog2000} and a brighter luminosity level by 0.25 dex in order to take into account a possible mild overshooting and/or mass loss efficiency. For each combination of chemical composition, mass and luminosity, we explore a wide range in effective temperature and derive the location of both the blue and red edge of the instability strip for each of the two investigated pulsation modes.\\

\begin{figure*}
\begin{center}
\includegraphics[width=1.0\textwidth,keepaspectratio]{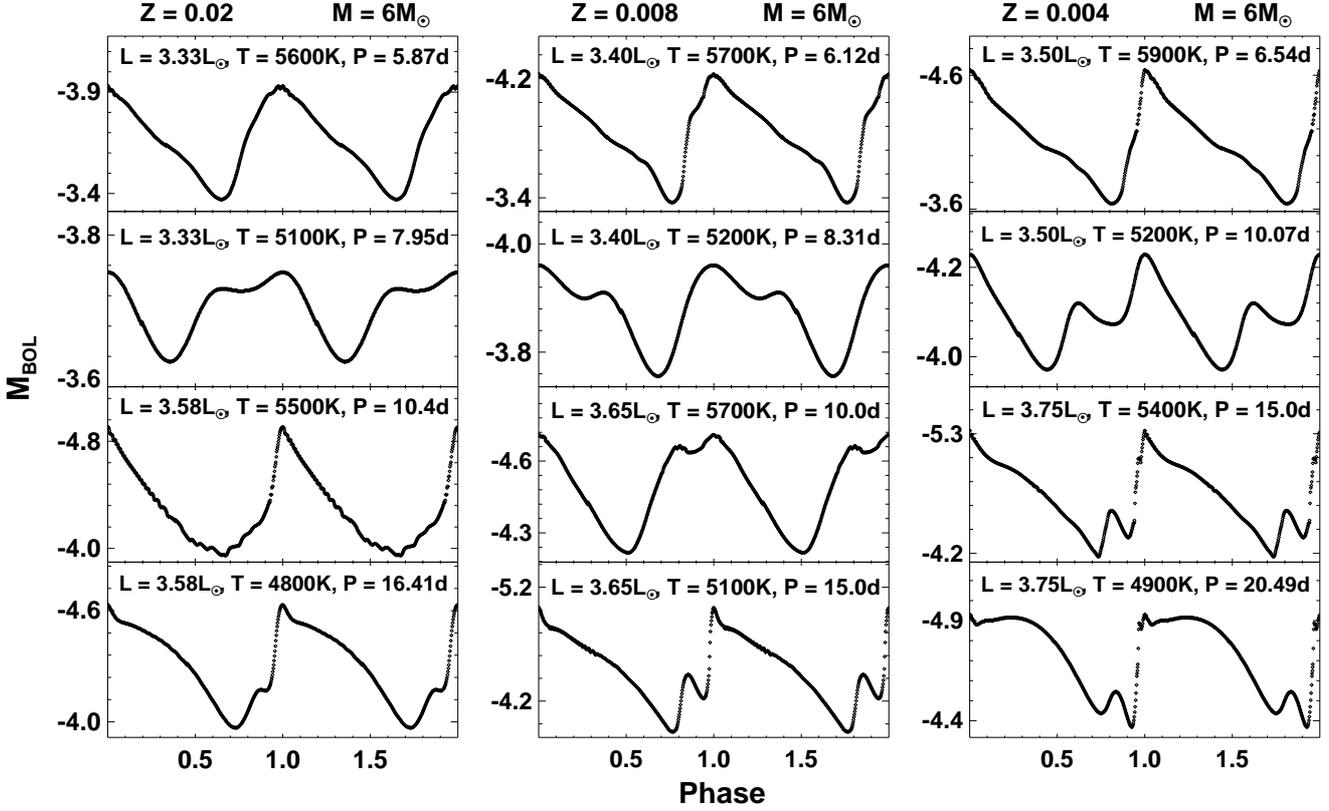}
\caption{Examples of theoretical light curves of Cepheid variables. Each column represents the predicted bolometric light curves for a fixed composition and mass with different luminosity levels and temperatures.}
\label{fig:lcs}
\end{center}
\end{figure*}

Fig.~\ref{fig:lcs} displays the predicted bolometric light curves for a fixed composition and mass with different luminosity levels and temperatures. The produced bolometric light curves are converted into the observational Johnson-Cousins optical ($UBVRI$) and NIR ($JKL$) filters by adopting the static model atmospheres by \citet{castelli1997a, castelli1997b}. These transformed curves, usually adopted to derive mean magnitude and colors, as well as pulsation amplitudes, can be decomposed through the Fourier analysis.\\

We also use the observed Cepheid light curve data for a comparison with theoretical light curves. We note that optical, NIR and mid-infrared (MIR) light-curve parameters of Cepheid variables in the Galaxy and LMC are taken from Tables 2 and 3 of \citet{bhardwaj2015a}. In this work, we update the optical sample with light curve data from OGLE-IV catalog \citep{ogle4cep} of Cepheid variables in the MC. The infrared data for Small Magellanic Cloud (SMC) Cepheids are taken from VMC survey \citep{ripepi2016} and \citet{scowcroft2016}. We carry out Fourier analysis of the $K_s$ and $3.6\mu\mathrm{m}$-band light curves in the SMC and present it here for the first time. Observed Cepheid data in $3.6\mu \mathrm{m}$-band will be used for relative comparison with theoretical light curves in $L$-band.

\begin{table*}
\begin{minipage}{1.0\hsize}
\begin{center}
\caption{The light curve parameters for the entire grid of models for Cepheid variables with composition relative to the Galaxy and MC. The first eight columns provide the chemical abundance, filter, stellar mass, pulsation mode, the effective temperature, the adopted logarithmic luminosity levels and the logarithmic period for each set of models and the last column represents the mean radius in the solar units. Other columns represent amplitude ($A$), Skewness ($S_k$), Acuteness ($A_c$) and Fourier parameters.}
\label{table:model_fou}
\begin{tabular}{cccccccccccccccc}
\hline
\hline
Z& Y& $\lambda$& $\frac{M}{M_\odot}$ & Mode & T$_{e}$& $\log \frac{L}{L_\odot}$ & $\log (P)$ & A& S$_k$& A$_c$& R$_{21}$& R$_{31}$& $\phi_{21}$& $\phi_{31}$& $\log \frac{R}{R_\odot}$  \\
\hline
0.020&0.28&U&5.4&FU&   5800&3.18&     0.620&      1.354&      2.215&      1.376&      0.296&      0.093&      2.734&      5.540&      1.586\\
0.020&0.28&U&5.4&FU&   5700&3.18&     0.645&      1.479&      2.390&      1.270&      0.322&      0.119&      2.817&      5.908&      1.601\\
0.020&0.28&U&5.4&FU&   5600&3.18&     0.670&      1.383&      2.333&      1.051&      0.321&      0.144&      2.910&      0.006&      1.616\\
0.020&0.28&U&5.4&FU&   5500&3.18&     0.696&      1.246&      2.205&      0.933&      0.319&      0.182&      3.021&      0.284&      1.632\\
0.020&0.28&U&5.4&FU&   5400&3.18&     0.723&      1.040&      2.135&      0.845&      0.330&      0.232&      3.167&      0.537&      1.648\\
0.020&0.28&U&5.4&FU&   5300&3.18&     0.746&      0.899&      3.587&      0.836&      0.354&      0.259&      3.271&      0.669&      1.665\\
0.020&0.28&U&5.4&FU&   5200&3.18&     0.776&      0.522&      2.425&      0.879&      0.331&      0.159&      3.205&      0.877&      1.681\\
0.020&0.28&U&5.4&FO&   5900&3.18&     0.429&      0.582&      1.463&      1.000&      0.160&      0.035&      3.138&      0.604&      1.573\\
0.020&0.28&U&5.4&FU&   5800&3.43&     0.837&      0.106&      1.151&      1.322&      0.139&      0.187&      0.073&      4.773&      1.711\\
0.020&0.28&U&5.4&FU&   5500&3.43&     0.916&      2.135&      0.992&      1.479&      0.121&      0.175&      0.131&      2.650&      1.757\\
\hline
\end{tabular}
\end{center}
{\footnotesize \textbf{Notes}: This table is available entirely in a machine-readable form in the online journal as supporting information.}
\end{minipage}
\end{table*}

\section{Fourier analysis}
\label{sec:fourier}

We apply the Fourier decomposition method as described in \citet{bhardwaj2015a}, to the theoretical light curves computed for a wide 
range of metallicities over multiple wavelengths. We use the Fourier sine series in the following form:

\begin{equation}
m = m_{0}+\sum_{k=1}^{N}A_{k} \sin(2 \pi k x + \phi_{k}),
\label{eq:foufit1}
\end{equation}

\noindent where, $x$ is the phase corresponding to one pulsation cycle of the theoretical light curves. The order of fit ($N$) is
obtained using the Baart's criteria \citep{baart82} by varying $N$ from 4-10. The Fourier coefficients ($A_{k}$ \& $\phi_k$)
are used to calculate the amplitude ratios and the phase differences ($R_{k1} = A_k/A_1~\&~ \phi_{k1} = \phi_k - k\phi_1, ~\mathrm{for}~ k > 1$). The errors on the Fourier coefficients resulting from the fit are only statistical, and are of the order of $10^{-4}$ and therefore, will not be considered further in our analysis. Table~\ref{table:model_fou} provides the light curve parameters for theoretical models of Cepheids with chemical compositions characteristic of the Galaxy, LMC and SMC. The $\phi_{31}$ values listed in Table~\ref{table:model_fou} are converted to cosine series by subtracting $\pi$ and restricting it between $0-2\pi$, for the continuity of HP for visualisation purposes. Although, at NIR bands this leads to a discontinuity in the progression, but we prefer to adopt cosine $\phi_{31}$ convention to ease the comparison with observed phase parameters from \citet{bhardwaj2015a}.

\begin{figure}
\begin{center}
\includegraphics[width=0.5\textwidth,keepaspectratio]{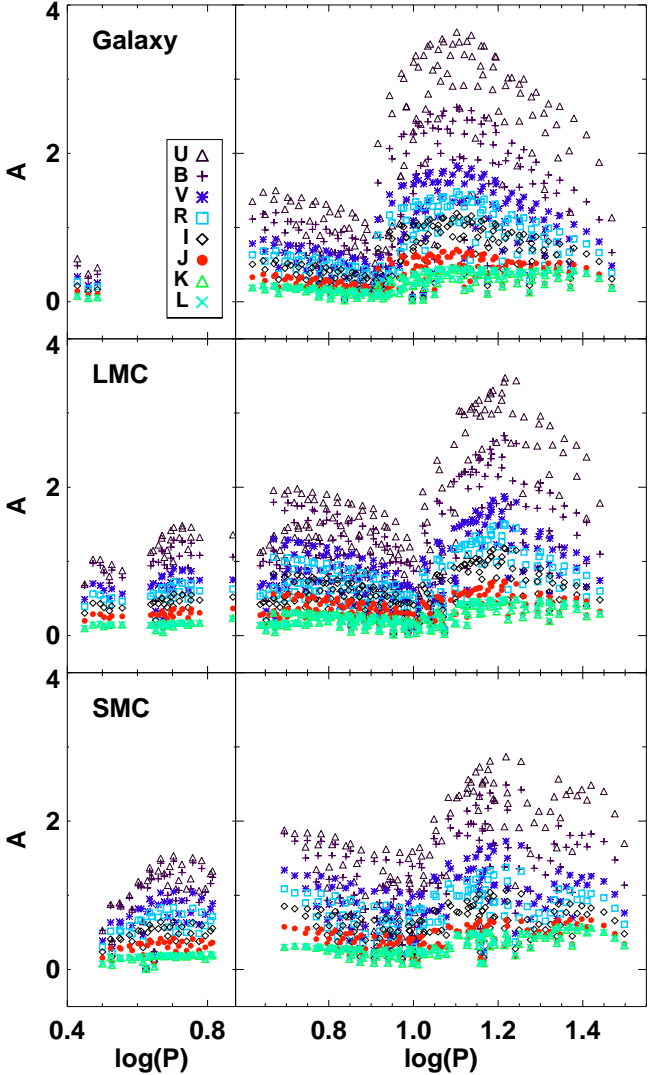}
\caption{The variation of the amplitude (A) as a function of period and wavelength for the theoretical light curves of Cepheids with metal abundance, Z=0.02, Z=0.008 and Z=0.004. The right/left panels of the figure display the FU/FO Cepheid models.}
\label{fig:galaxy_amp}
\end{center}
\end{figure}
\begin{figure}
\begin{center}
\includegraphics[width=0.5\textwidth,keepaspectratio]{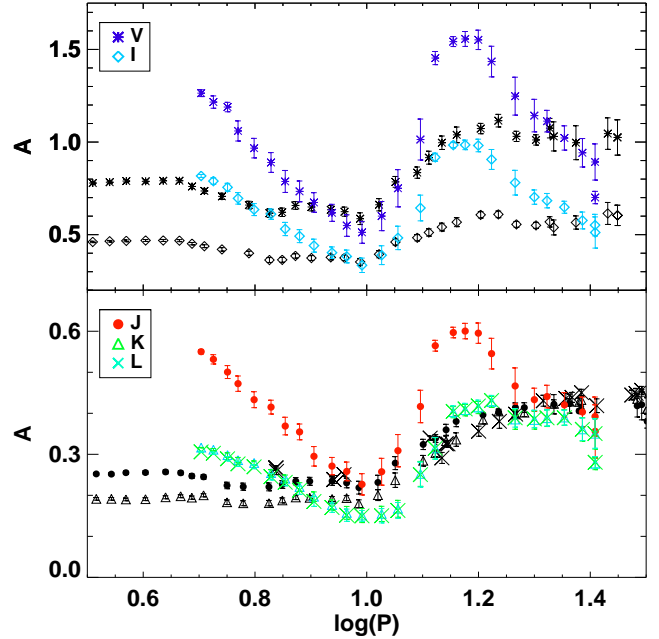}
\caption{A comparison of observed (black symbols) and theoretical (colored symbols) mean amplitudes for FU Cepheids in the LMC.}
\label{fig:amp_lmc}
\end{center}
\end{figure}

We study the variation in amplitude as a function of period, wavelength and metallicity. The amplitude is defined as the difference of maximum and minimum magnitudes obtained from the best order Fourier fit. Fig.~\ref{fig:galaxy_amp} displays the amplitudes for FU and FO Cepheids with chemical compositions representative of the Galaxy and MC. We observe a decrease in the peak-to-peak amplitude with increasing wavelength for both FU and FO Cepheids. We find a sharp increase in the amplitude parameters in the period range, $1 < \log (P) < 1.2$, for optical bands as compared to infrared bands, similar to observed amplitudes in \citet{bhardwaj2015a}. For $\log (P) > 1.2$, the amplitudes decrease with period at all wavelengths. For SMC composition, a secondary maximum occurs at $\log (P) = 1.4$ for FU Cepheid models.

Fig.~\ref{fig:amp_lmc} shows a comparison of observed and theoretical mean amplitudes for Cepheids in the LMC. The mean values are obtained using a bin-size of $\log (P)=0.1$~dex and in steps of $\sim0.03$~dex. The error bars represent the standard deviation on the mean in each bin. We note that the optical ($VI$) and $J$ band mean amplitudes are systematically larger than observed amplitudes as a function of period, except in the vicinity of 10 days. The observed mean amplitudes range, $0.1 < A \lesssim 1.2$~mag, from infrared (bottom panel) to optical ($VI$, top panel) bands while the theoretical amplitudes have a greater range upto $\sim1.6$~mag, for this wavelength range. The $K$- and $L$-band amplitudes are consistent between theory and observations in most period bins. \citet{bono2002} suggested that the discrepancy in theoretical amplitudes can be decreased with increase in mixing length parameter ($\alpha$). The impact of increase in mixing length on the light curve parameters will be discussed in the last section.

\subsection{Theoretical Fourier parameters}

\subsubsection{As a function of stellar mass}

\begin{figure*}
\begin{center}
\includegraphics[width=1.0\textwidth,keepaspectratio]{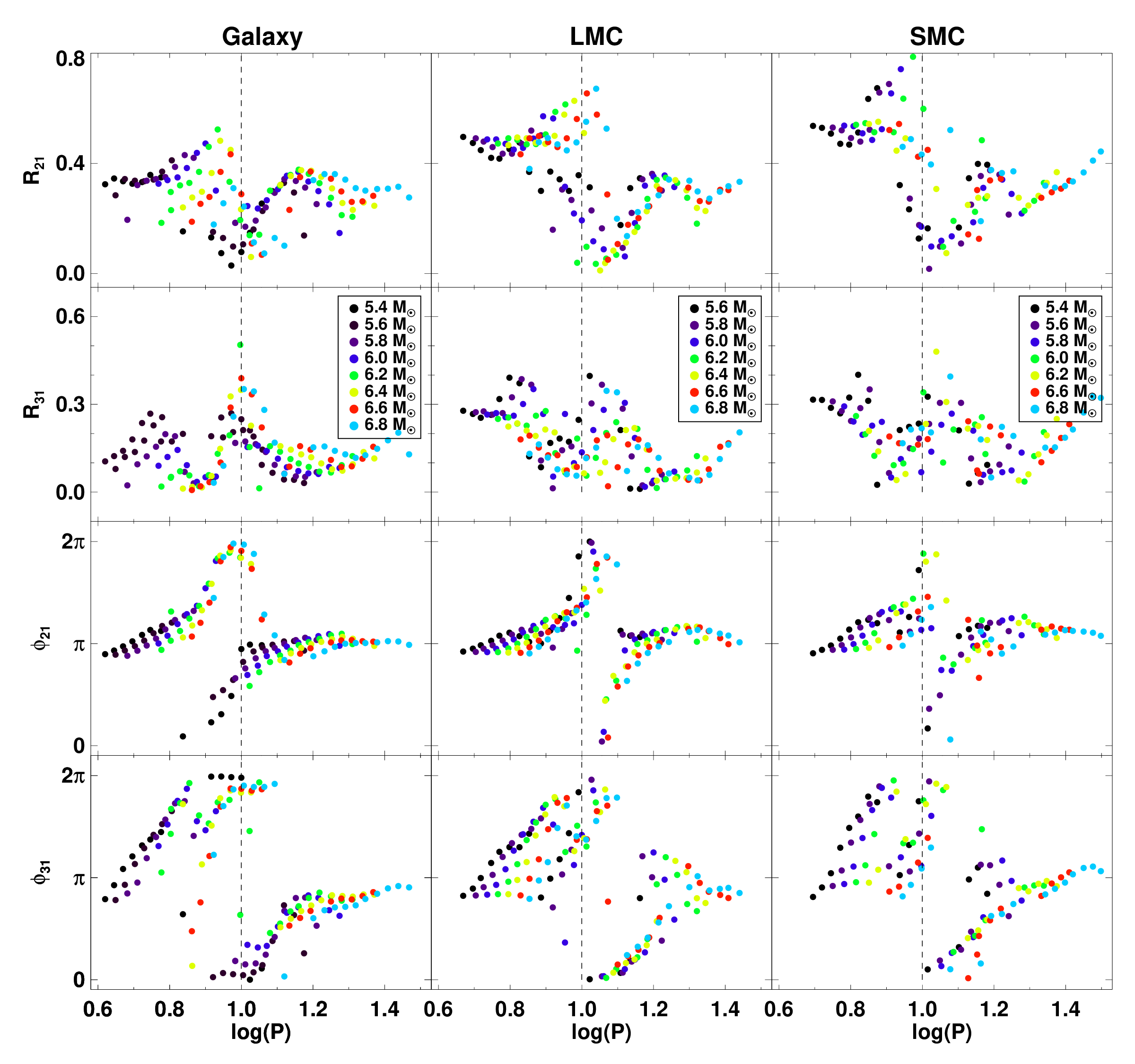}
\caption{Fourier parameters for theoretical light curves of Cepheids in the $V$-band with chemical composition relative to the Galaxy, LMC and SMC. The vertical dashed lines represent 10 day period as expected center of HP.}
\label{fig:ml_v}
\end{center}
\end{figure*}

\begin{figure*}
\begin{center}
\includegraphics[width=1.0\textwidth,keepaspectratio]{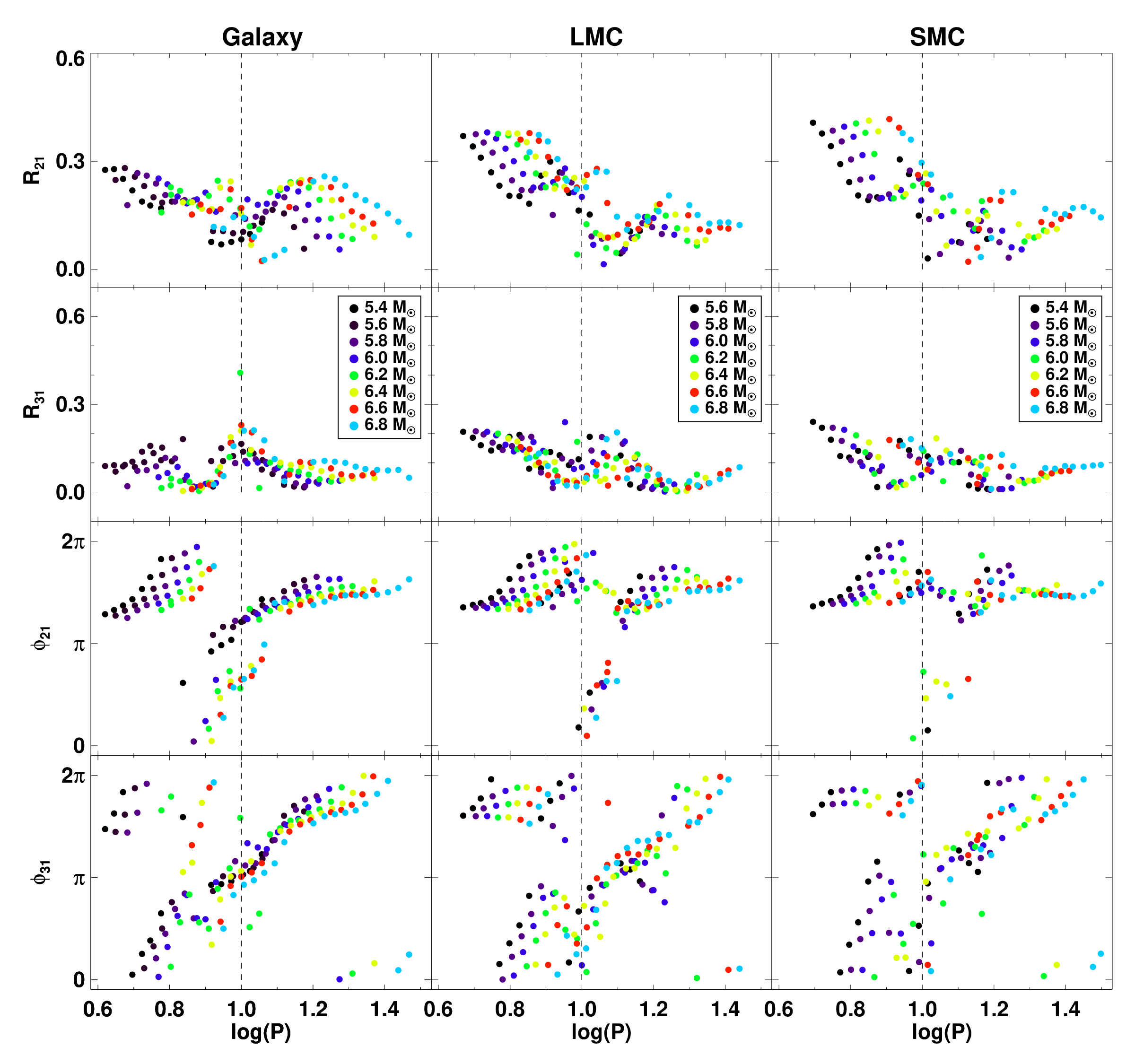}
\caption{Same as Fig.~\ref{fig:ml_v} but for $K$-band.}
\label{fig:ml_k}
\end{center}
\end{figure*}

\begin{figure*}
\begin{center}
\includegraphics[width=1.0\textwidth,keepaspectratio]{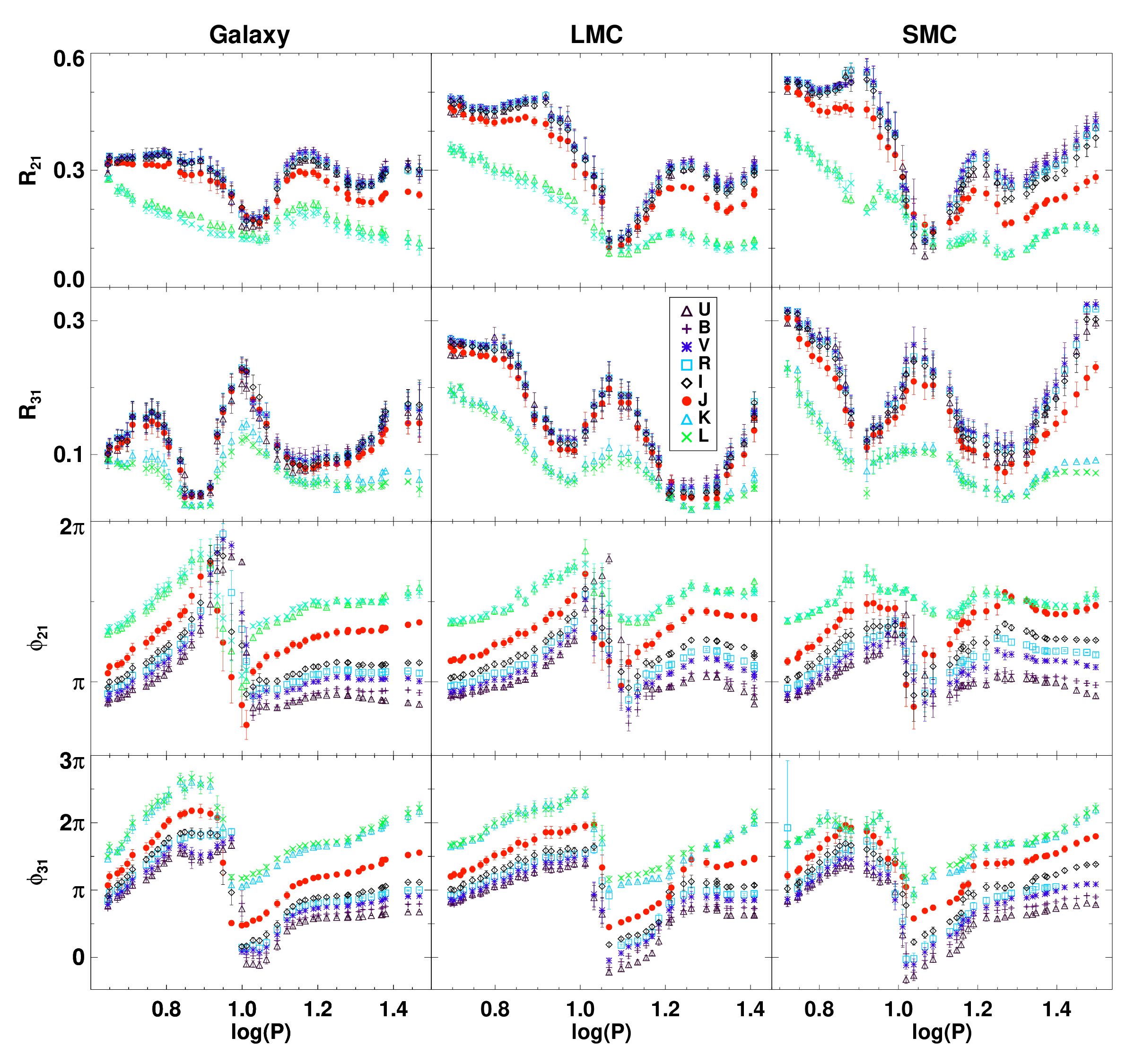}
\caption{Multi-wavelength mean Fourier parameters for the theoretical light curves of Cepheids with chemical compositions relative to the Galaxy, LMC and SMC. The error bars represent the standard deviation on the mean. Some of the cosine $\phi_{31}$ parameters are offset by $2\pi$, and excluded from 0-$2\pi$ restriction, only for plotting purposes.}
\label{fig:mean_fou}
\end{center}
\end{figure*}

\begin{figure*}
\begin{center}
\includegraphics[width=1.0\textwidth,keepaspectratio]{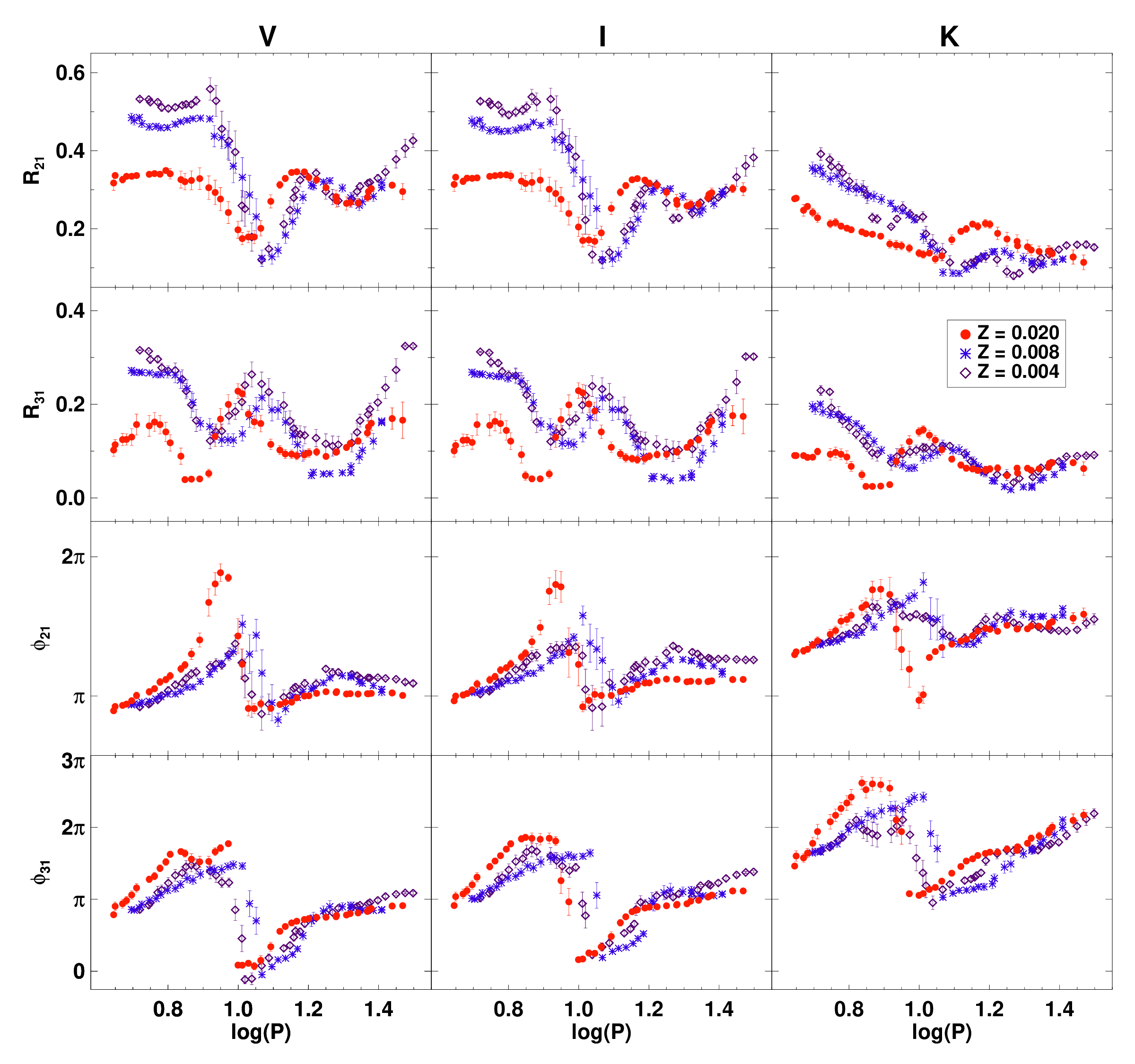}
\caption{A comparison of the theoretical mean Fourier parameters for FU Cepheids with different compositions.}
\label{fig:fou_metal}
\end{center}
\end{figure*}


\begin{figure*}
\begin{center}
\includegraphics[width=1.\textwidth,keepaspectratio]{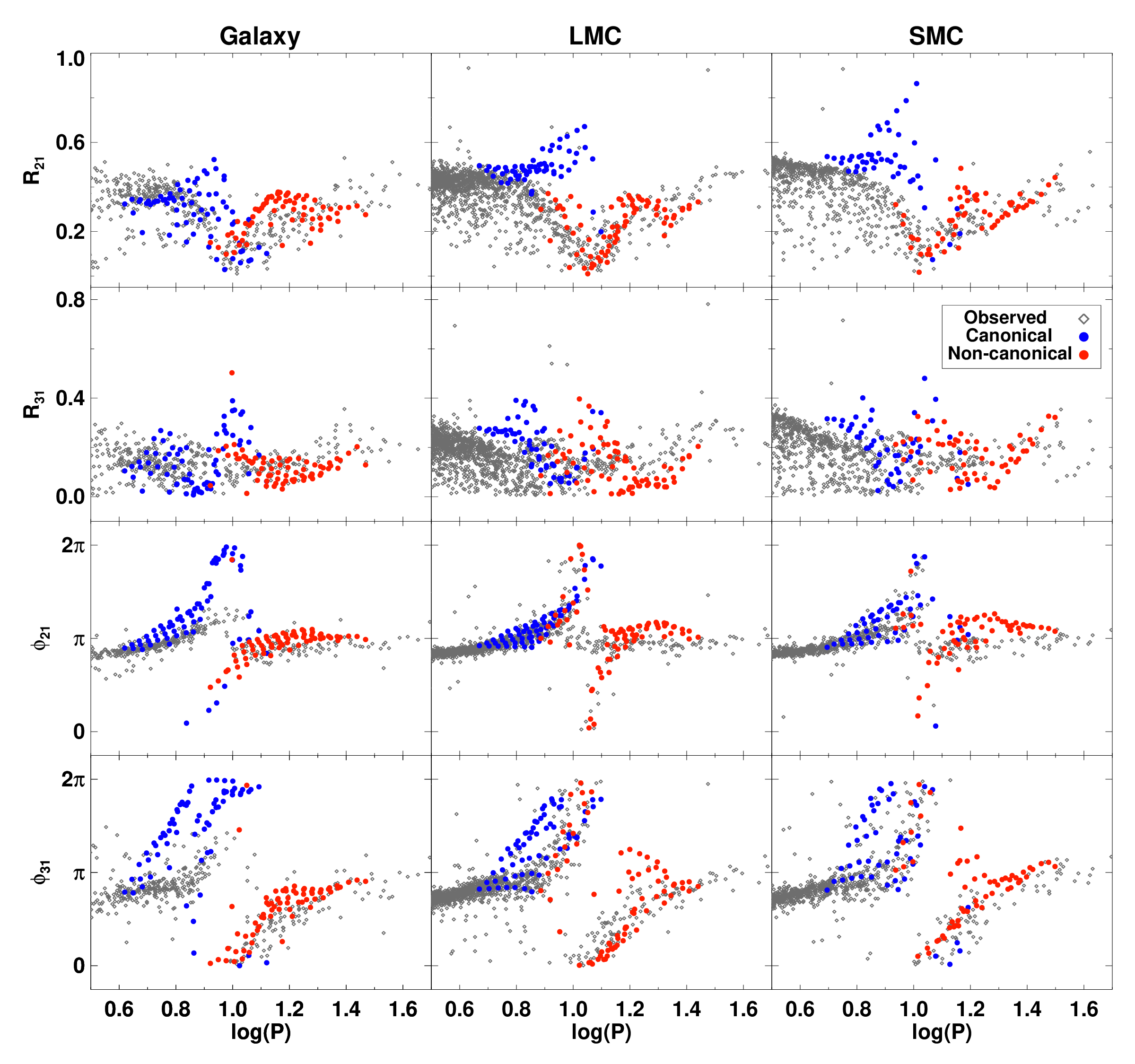}
\caption{A comparison of the theoretical and observed $V$-band Fourier parameters for FU Cepheids in the Galaxy, LMC and SMC. Canonical represents luminosity levels adopted from stellar evolutionary calculations while non-canonical represents brighter luminosity levels by 0.25 dex.}
\label{fig:theory_ogle_v}
\end{center}
\end{figure*}

\begin{figure*}
\begin{center}
\includegraphics[width=1.0\textwidth,keepaspectratio]{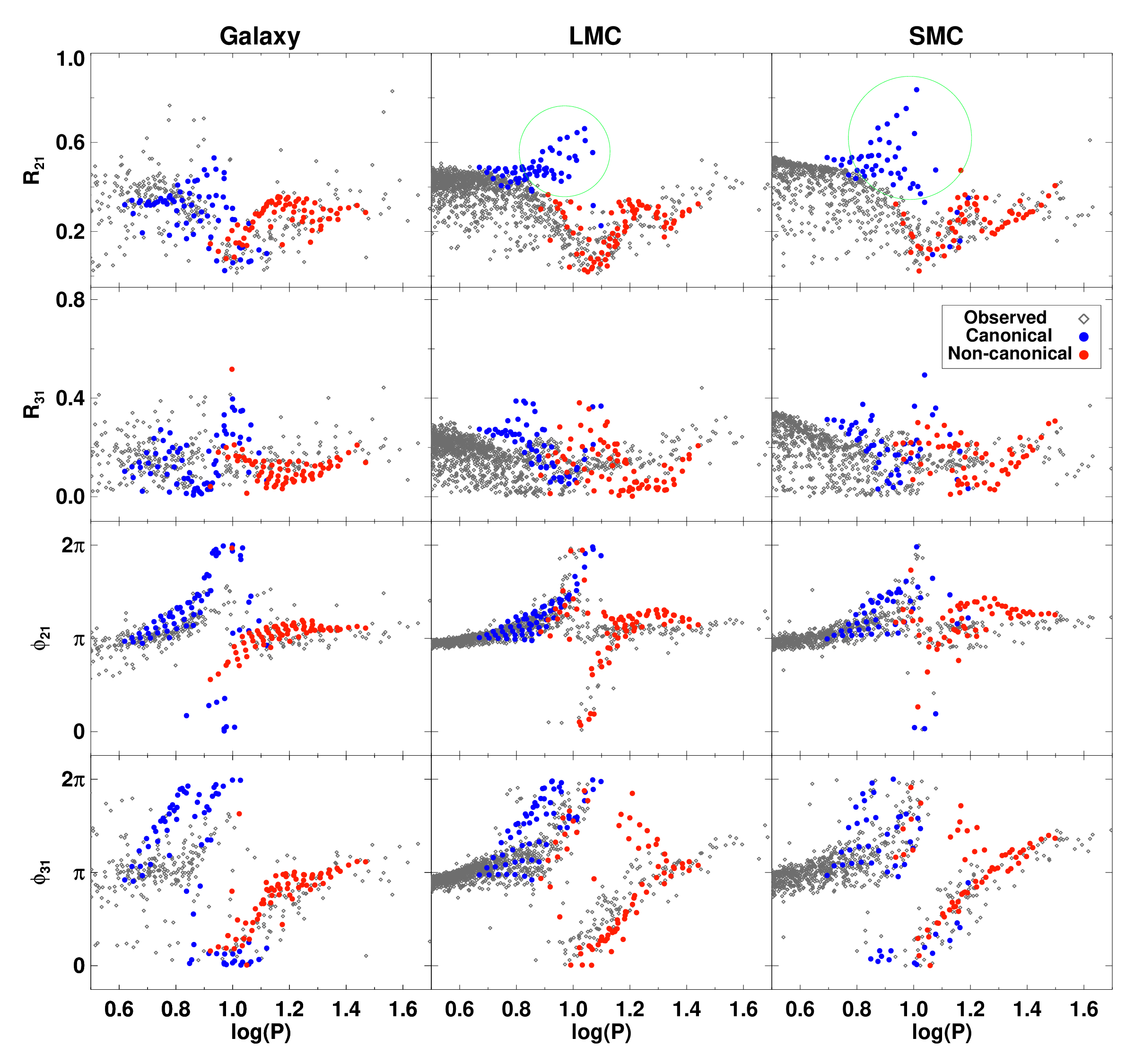}
\caption{Same as Fig.~\ref{fig:theory_ogle_v} but for $I$-band Fourier parameters. The points within the green circle in $R_{21}$ are selected as subset of models showing large offset with respect to observations.}
\label{fig:theory_ogle_i}
\end{center}
\end{figure*}

\begin{figure*}
\begin{center}
\includegraphics[width=1.\textwidth,keepaspectratio]{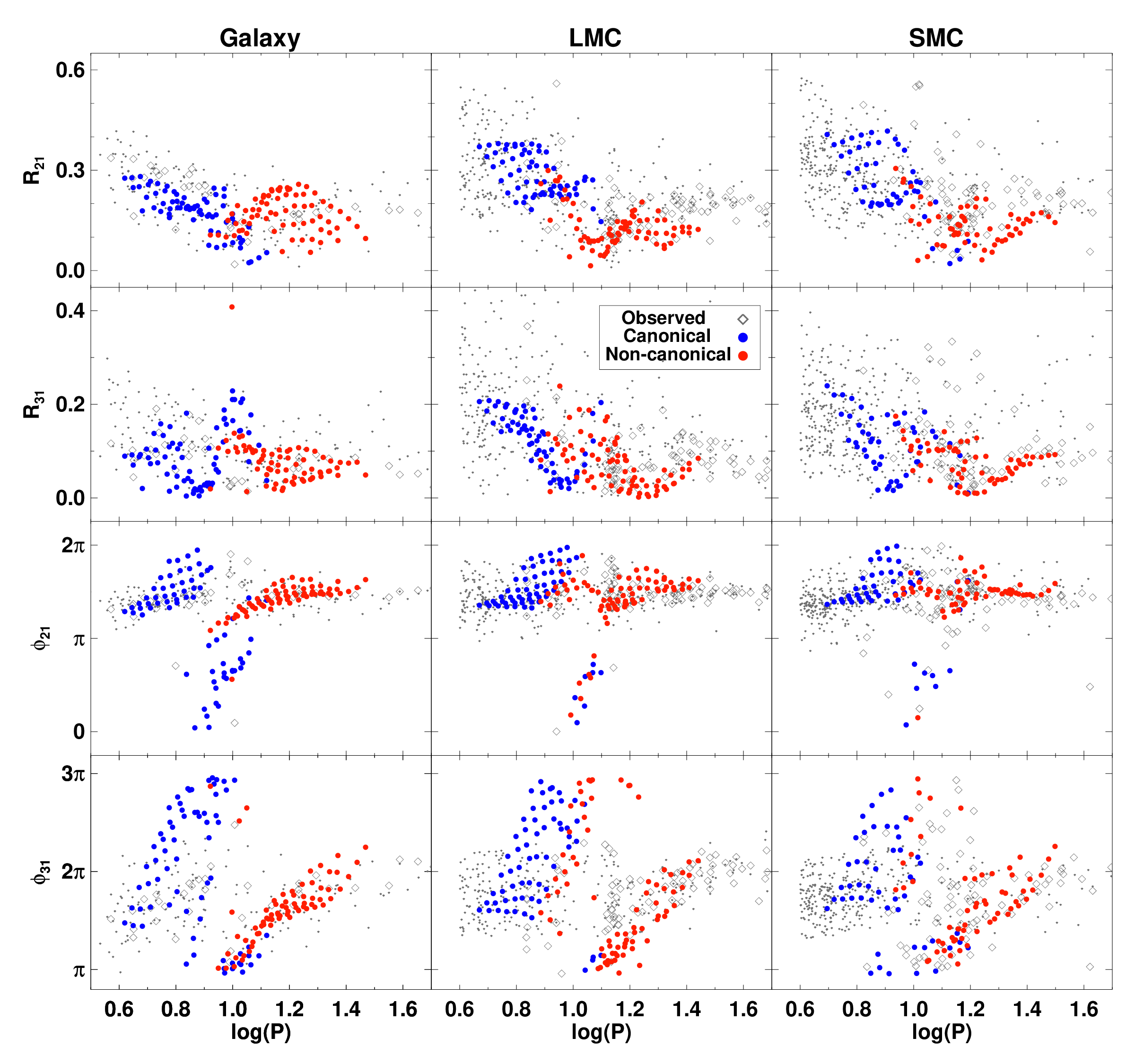}
\caption{Same as Fig.~\ref{fig:theory_ogle_v} but for $K$-band (dots) and $3.6\mu \mathrm{m}$-band (diamonds) observations compared to $K$-band canonical/non-canonical models.}
\label{fig:theory_ogle_k}
\end{center}
\end{figure*}


We present the $V$ and $K$-band Fourier parameters for theoretical light curves with different chemical compositions (Z=0.02, Z=0.008 and Z=0.004) in Fig.~\ref{fig:ml_v} and \ref{fig:ml_k}, respectively. Theoretical Fourier parameters in other optical and NIR bands show similar trends to $V$ and $K$-bands, respectively, as a function of ML levels. The HP is clearly observed for $R_{21}$ in the vicinity of $\log (P)=1.0$ for $UBVRIJKL$ wavelengths while the primary minimum in $R_{31}$ occurs around $\log (P)=0.9$. These amplitude parameters show a sharp rise around $1.0<\log (P)<1.2$ in optical bands. The $\phi_{21}$ and $\phi_{31}$ parameters also suggest the center of HP around a period of 10 days. The scatter in amplitude parameters is least in NIR bands and increases for shorter wavelengths. 

We show the different stellar mass models with different colors in these plots. For galactic models, we find that the amplitude parameters for short and long period Cepheids show different progression for a fixed mass at optical bands. The amplitude parameters decrease with increasing mass at short period end, but increase with mass at the long period end. However, no distinct trend is observed for period range close to the center of HP. Furthermore, $R_{21}$ and $R_{31}$ display a larger scatter in optical for lower metallicity models. The scatter reduces in NIR but no distinct variation with adopted ML levels is seen, presumably due to small amplitudes. Moreover, larger mass leads to a decrease in the phase parameters at a given period for all wavelengths and metallicities, specifically in $\phi_{21}$ parameters. The progression in $\phi_{31}$ is more evident in the optical bands, due to break in the HP at NIR wavelengths, on account of $0-2\pi$ restriction on phase parameters. We also note a secondary minimum in the $R_{21}$ parameters around $\log (P) \sim 1.3$ which shifts to earlier periods in case of $R_{31}$ and is more pronounced for lower metallicity models.

\subsubsection{As a function of wavelength}

We over-plot the Fourier parameters for optical and infrared bands with different colors and symbols in Figs.~\ref{fig:z02y28_multiband} and ~\ref{fig:mc_multiband} for Galaxy and MC Cepheid models, respectively, to compare directly the variation as a function of wavelength. 

Following \citet{bhardwaj2015a}, we use sliding mean calculations to examine the dependence of the Fourier parameters on wavelength and metallicity, simultaneously. Since the theoretical Fourier parameters do not have errors on the individual data points and exhibit less scatter, we use reduced step size of $\log (P)=0.02$~dex and a bin width of 0.1~dex to estimate moving averages. We also estimate the standard deviation of the mean in each step as a representative statistical error on the theoretical mean Fourier parameters.

Fig.~\ref{fig:mean_fou} displays the variation of the mean Fourier parameters as a function of period and wavelength for metallicity, Z=0.02, Z=0.008 and Z=0.004. We observe a decrease in amplitude parameters and an increase in phase parameters as a function of wavelength. The amplitude parameters systematically increase from Galaxy to MC for short period Cepheid models. For longer period Cepheids, $R_{21}$ increases sharply from $\log (P)=1$ to reach a peak value around $\log (P)=1.2$ before secondary minimum occurs around $\log (P)=1.3$. This secondary feature is more pronounced for lower metallicity models and, at optical wavelengths than for the redder bands. Similar separation is also seen in the observed amplitude parameters in the Galaxy and the LMC around 20 days \citep{bhardwaj2015a}. This may suggest a possible resonance at $\log (P)=1.3$ and these changes in the light curve structure can be related with observed non-linearity in P-L relations at similar periods in \citet{cpapir2, cpapir3}. The largest separation for $R_{21}$ occurs between optical and NIR bands around $\log (P)=0.8$ for the Galaxy and $\log (P)=0.9$ for the MC models. The $R_{31}$ shows a minimum before and after 10 days with a peak value at $\log (P)=1.0$ for the Galaxy and closer to $\log (P)=1.1$ for the MC. However, no significant difference is seen in $R_{31}$ for wavelengths shorter and longer than $J$-band, around $\log (P)=1.2$, as observed for $R_{21}$. The mean phase parameters exhibit a systematic separation at all wavelengths for the Galaxy and MC models and the center of HP occurs in the vicinity of 10 days. The mean $\phi_{21}$ parameters for Galactic light curves are nearly constant in optical bands for $\log (P) > 1.2$.

\subsubsection{As a function of metal abundance}

In Fig.~\ref{fig:fou_metal}, we compare the theoretical mean Fourier parameters for FU Cepheids as a function of metal abundance in $VIK$ bands. We find that the metal-rich Galactic models have significantly different Fourier parameters compared to metal-poor MC Cepheid models in most period bins. For $\log (P) < 1.0$, there is a distinct trend in amplitude parameters with $R_{21}$ and $R_{31}$ increasing with decreasing metallicity. In case of the phase parameters, the value of $\phi_{21}$ decreases with metallicity for $\log (P) < 1.0$ while it increases with decreasing metallicity for $\log (P) > 1.1$. The progression is not clear close to the center of HP and for low metallicity MC Cepheid models. The $\phi_{31}$ parameters also provide evidence of a decrease with metal abundance for most period ranges, except for $\log (P) > 1.2$.

It is also evident that the center of HP occurs at longer periods with decreasing metallicity. We fit a polynomial through mean $R_{21}$ in the period range $0.8 < \log (P) < 1.2$ and obtain the minimum value of the functional fit. The center of HP is found to be at $\log (P) = 1.030\pm0.011$~dex for the Galaxy, $\log (P) = 1.092\pm0.015$~dex for the LMC and at $\log (P) = 1.077\pm0.010$~dex for the SMC at optical ($VI$) bands. In case of $K$-band, the central period shifts to $\log (P) = 1.046\pm0.009$ for the Galaxy, $\log (P) = 1.103\pm0.016$ for the LMC and at $\log (P) = 1.108\pm0.014$ for the SMC. The difference in central period for MC Cepheid models is not significant, given their statistical uncertainties. The central period of HP for the Galaxy is in excellent agreement with observed value using optical $R_{21}$ in \citet{bhardwaj2015a}. Similarly, the value of the central period for the LMC is consistent within $3\sigma$ with observations in $VIK$ bands. This confirms the shift in central period of HP to longer periods with lower metallicity, and also with longer wavelengths in case of $R_{21}$ \citep{bhardwaj2015a}. We do not observe any significant trend with metallicity in amplitude parameters for Cepheids having periods greater than 10 days.

\subsection{Comparison of theoretical and observed Fourier parameters}

\begin{figure*}
\begin{center}
\includegraphics[width=1.0\textwidth,keepaspectratio]{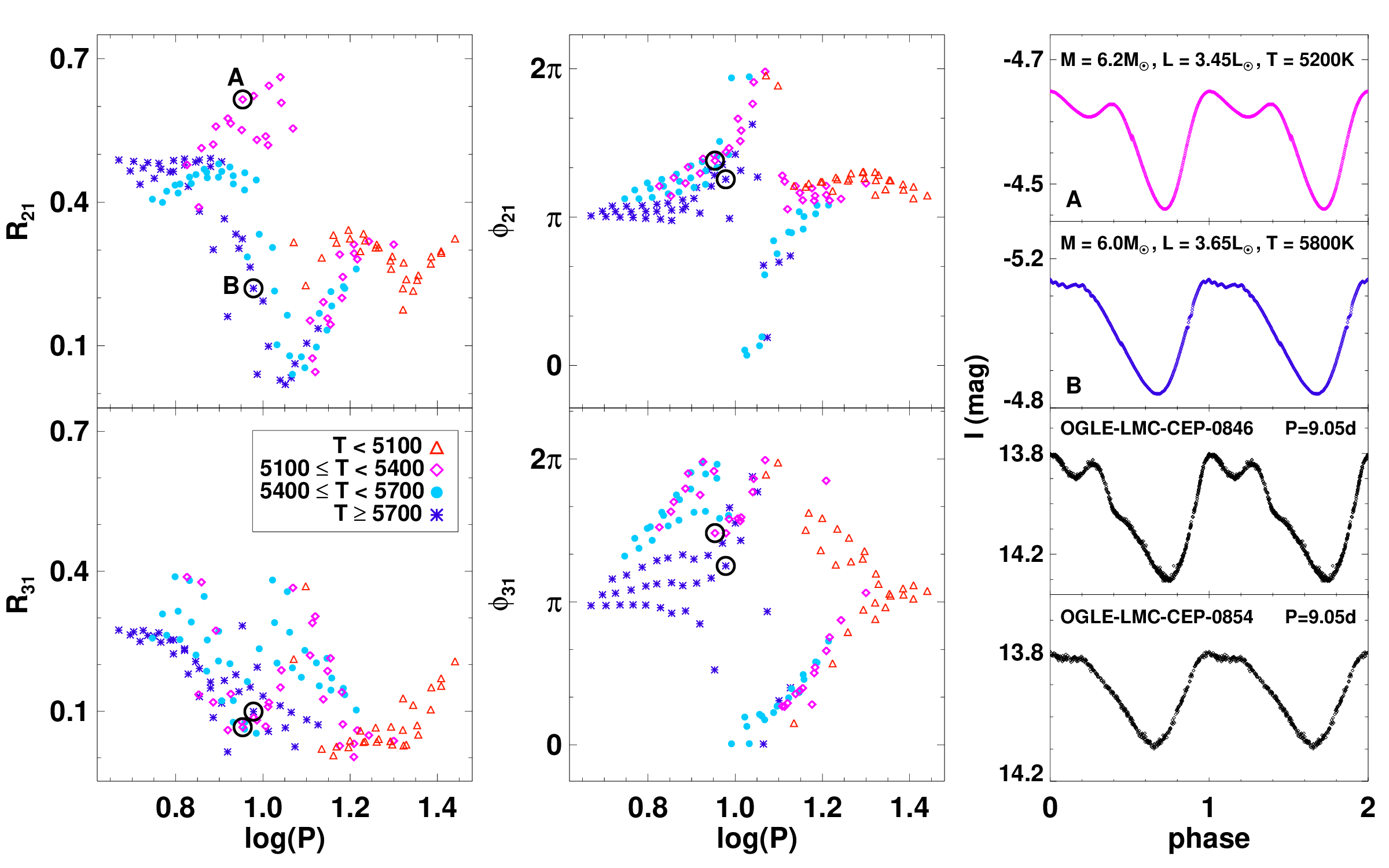}
\caption{The variation of $I$-band Fourier parameters with temperature at a given period for LMC Cepheid models. The selected models (within black circles) in $R_{21}$ plane represent the different parameter values with similar periods and the top right panel displays the difference in their light curves and adopted ML relations. The bottom right panels show the observed light curves with similar period and $R_{21}$ values as that of the selected pair of Cepheid models.}
\label{fig:temp_fou}
\end{center}
\end{figure*}

We compare the Fourier parameters of the theoretical $VIK$-band light curves with observed $VIK$-band Fourier parameters for the Cepheid light curves in the Galaxy and MC. The comparison of the optical band Fourier parameters is more rigourous as the observed NIR light curves are not as well sampled as their optical counterparts. The observed optical $V$ and $I$ band Fourier parameters are very accurate and exhibit least error \citep{bhardwaj2015a}.

Figs.~\ref{fig:theory_ogle_v} and \ref{fig:theory_ogle_i} display the Fourier parameters of the theoretical and observed light curves of FU Cepheids in the Galaxy and MC in $V$ and $I$-band, respectively. We also plot the canonical and non-canonical ML models in different colors. For Galactic data, the theoretical amplitude parameters follow the observed progressions with a marginal offset at the long period end ($\log (P) > 1.1$). In case of $\phi_{21}$, the theoretical and observed progressions are consistent while theoretical $\phi_{31}$ are significantly larger for short period Cepheids ($\log (P) < 0.9$). In case of the MC, $R_{21}$ and $R_{31}$ are systematically larger in theory at the short period end ($\log (P) < 1$). It is evident that the non-canonical ML models follow the observations while the canonical models are inconsistent in the overlapping period range ($0.8 < \log (P) < 1.1$). Therefore, $R_{21}$ plane does seem to differentiate between the canonical and non-canonical ML models. Furthermore, the offset between theory and observations increases with decreasing metallicity in this period range. For $\log (P) > 1.1$, there is an agreement for $R_{21}$ between theory and observations. However, $R_{31}$ is inconsistent in most period ranges and the central minimum is not well defined for both theoretical models and observations. Similarly, there is a small offset for $\phi_{21}$ and $\phi_{31}$ at the long period end ($\log (P) > 1.2$). At present, we have a lack of models for FU Cepheids with $\log (P) < 0.6$. 

Fig.~\ref{fig:theory_ogle_k} displays the Fourier parameters of the theoretical and observed light curves of FU Cepheids in the Galaxy, LMC and SMC in $K$-band. Since, the $K$ and $L$-band theoretical light curves exhibit very similar Fourier parameters, we also include the more precise $3.6\mu \mathrm{m}$-band Fourier parameters for Cepheids in the Galaxy and MC. We do not find any large offset in amplitude parameters with respect to observation at this wavelength. However, long-period ($\log (P) > 1.2$) Cepheid models do show an offset for lower metallicity models but it is not conclusive, given the large scatter in Fourier parameters at longer wavelengths. The $\phi_{21}$ parameters are consistent over entire period range and the $\phi_{31}$ parameter shows a greater offset for short-period range Cepheid models, similar to optical-band data.

\begin{table}
\begin{minipage}{1.0\hsize}
\begin{center}
\caption{The model parameters for MC Cepheid models which do not match with observations on the Fourier $R_{21}$ plane at optical wavelengths. \label{table:diff_fou}}
\begin{tabular}{ccp{0.7cm}cccp{0.7cm}c}
\hline
\hline
Z&  $\frac{M}{M_\odot}$ & $\log \frac{L}{L_\odot}$ & $\log (P)$ & $M_V$ & T$_{e}$& $\log \frac{R}{R_\odot}$ & $R_{21}$  \\
\hline
0.008&5.8&3.35&     0.859&     -3.476&    5300&      1.750&      0.513\\
0.008&5.8&3.35&     0.886&     -3.448&    5200&      1.767&      0.521\\
0.008&6.0&3.40&     0.865&     -3.628&    5400&      1.759&      0.469\\
0.008&6.0&3.40&     0.892&     -3.602&    5300&      1.775&      0.558\\
0.008&6.0&3.40&     0.920&     -3.573&    5200&      1.792&      0.575\\
0.008&6.2&3.45&     0.846&     -3.797&    5600&      1.752&      0.458\\
0.008&6.2&3.45&     0.872&     -3.776&    5500&      1.768&      0.451\\
0.008&6.2&3.45&     0.899&     -3.753&    5400&      1.784&      0.480\\
0.008&6.2&3.45&     0.926&     -3.727&    5300&      1.800&      0.565\\
0.008&6.2&3.45&     0.954&     -3.699&    5200&      1.817&      0.615\\
0.008&6.4&3.49&     0.846&     -3.917&    5700&      1.756&      0.484\\
0.008&6.4&3.49&     0.871&     -3.898&    5600&      1.772&      0.462\\
0.008&6.4&3.49&     0.898&     -3.877&    5500&      1.788&      0.452\\
0.008&6.4&3.49&     0.924&     -3.853&    5400&      1.804&      0.474\\
0.008&6.4&3.49&     0.951&     -3.827&    5300&      1.820&      0.551\\
0.008&6.4&3.49&     0.979&     -3.799&    5200&      1.837&      0.622\\
0.008&6.4&3.49&     1.006&     -3.767&    5100&      1.854&      0.538\\
0.008&6.6&3.54&     0.854&     -4.061&    5800&      1.766&      0.488\\
0.008&6.6&3.54&     0.880&     -4.043&    5700&      1.781&      0.492\\
0.008&6.6&3.54&     0.906&     -4.024&    5600&      1.797&      0.464\\
0.008&6.6&3.54&     0.932&     -4.003&    5500&      1.813&      0.438\\
0.008&6.6&3.54&     0.958&     -3.979&    5400&      1.829&      0.462\\
0.008&6.6&3.54&     0.986&     -3.953&    5300&      1.845&      0.531\\
0.008&6.6&3.54&     1.014&     -3.924&    5200&      1.862&      0.644\\
0.008&6.6&3.54&     1.042&     -3.892&    5100&      1.879&      0.608\\
0.008&6.8&3.58&     0.880&     -4.162&    5800&      1.786&      0.475\\
0.008&6.8&3.58&     0.905&     -4.144&    5700&      1.801&      0.485\\
0.008&6.8&3.58&     0.932&     -4.124&    5600&      1.817&      0.455\\
0.008&6.8&3.58&     0.958&     -4.103&    5500&      1.833&      0.425\\
0.008&6.8&3.58&     0.985&     -4.079&    5400&      1.849&      0.446\\
0.008&6.8&3.58&     1.012&     -4.053&    5300&      1.865&      0.519\\
0.008&6.8&3.58&     1.040&     -4.024&    5200&      1.882&      0.662\\
0.008&6.8&3.58&     1.069&     -3.992&    5100&      1.899&      0.554\\
0.004&5.4&3.35&     0.849&     -3.487&    5400&      1.735&      0.599\\
0.004&5.4&3.35&     0.874&     -3.465&    5300&      1.750&      0.665\\
0.004&5.6&3.40&     0.853&     -3.635&    5500&      1.744&      0.478\\
0.004&5.6&3.40&     0.880&     -3.612&    5400&      1.760&      0.612\\
0.004&5.6&3.40&     0.907&     -3.588&    5300&      1.777&      0.683\\
0.004&5.8&3.45&     0.833&     -3.799&    5700&      1.738&      0.492\\
0.004&5.8&3.45&     0.887&     -3.760&    5500&      1.769&      0.474\\
0.004&5.8&3.45&     0.913&     -3.738&    5400&      1.785&      0.599\\
0.004&5.8&3.45&     0.940&     -3.713&    5300&      1.802&      0.721\\
0.004&6.0&3.50&     0.815&     -3.958&    5900&      1.733&      0.536\\
0.004&6.0&3.50&     0.841&     -3.943&    5800&      1.748&      0.537\\
0.004&6.0&3.50&     0.866&     -3.925&    5700&      1.763&      0.495\\
0.004&6.0&3.50&     0.920&     -3.886&    5500&      1.794&      0.461\\
0.004&6.0&3.50&     0.947&     -3.863&    5400&      1.810&      0.574\\
0.004&6.0&3.50&     0.974&     -3.838&    5300&      1.827&      0.752\\
0.004&6.0&3.50&     1.003&     -3.811&    5200&      1.843&      0.640\\
0.004&6.0&3.75&     1.166&     -4.412&    5100&      1.983&      0.474\\
0.004&6.2&3.55&     0.851&     -4.086&    5900&      1.757&      0.541\\
0.004&6.2&3.55&     0.877&     -4.070&    5800&      1.772&      0.542\\
0.004&6.2&3.55&     0.928&     -4.033&    5600&      1.803&      0.410\\
0.004&6.2&3.55&     0.953&     -4.014&    5500&      1.818&      0.440\\
0.004&6.2&3.55&     1.010&     -3.965&    5300&      1.851&      0.836\\
0.004&6.6&3.64&     0.908&     -4.311&    5900&      1.804&      0.521\\
0.004&6.6&3.64&     0.935&     -4.297&    5800&      1.817&      0.535\\
0.004&6.8&3.69&     0.945&     -4.439&    5900&      1.827&      0.461\\
0.004&6.8&3.69&     0.968&     -4.421&    5800&      1.844&      0.479\\
0.004&6.8&3.69&     0.998&     -4.406&    5700&      1.857&      0.414\\
0.004&6.8&3.69&     1.078&     -4.343&    5400&      1.904&      0.476\\
\hline
\end{tabular}
\end{center}
\end{minipage}
\end{table}

Most of the theoretical models with $M\geq6.0M_\odot$ are responsible for the large $R_{21}$ values for $0.8 < \log (P) < 1.1$ in the LMC (see, Fig.~\ref{fig:theory_ogle_i}). These models differentiate the canonical set of models from the non-canonical in the overlapping period range. However, we note that these outliers in the $R_{21}$ plane are reasonably consistent with observations in case of other Fourier parameters. We investigate the possible cause of discrepancy for this particular group of models having large offsets. Their parameters are listed in Table~\ref{table:diff_fou} for LMC and SMC. In Fig.~\ref{fig:temp_fou}, we plot the variation of Fourier parameters with temperature and period for the Cepheid models in the LMC. We find that the models which are outliers in $R_{21}$, with $\log (P) < 1.1$, have relatively lower temperatures, $5100 \leq T < 5400$ as compared to non-canonical Cepheid models. We select a pair of models in $R_{21}$ which have similar periods but different parameter values. It is evident that other Fourier parameters for these two models are consistent and the discrepancy is only in $R_{21}$. The light curves of these models are shown in the top right panel of Fig.~\ref{fig:temp_fou}. The light curves of the models with a large offset in $R_{21}$ plane show a bump around the maximum value. This suggests that the theoretical models produce bump Cepheid progressions but with a larger secondary amplitude when compared to the observations. We note that the pair of light curves has different adopted ML relations and effective temperatures. We also display two observed light curves with similar period and $R_{21}$ values on the bottom right panels of Fig.~\ref{fig:temp_fou}. We find that the light curve structure in theory and observations show similar features around maximum light. This can also allow us to provide an initial estimate for the ML levels and temperatures for the observed Cepheids and will be investigated in a future work through machine-learning methods.

\section{Principal Component Analysis}
\label{sec:pca}

\begin{table}
\begin{center}
\caption{The first 10 eigenvalues ($\lambda$) and their cumulative percentage (C.P.) of variance resulting from PCA analysis to :
(A) multiwavelength theoretical light curves of Cepheids with composition, Z=0.02 Z=0.008 and Z=0.004 
(B) theoretical and observed light curves at optical wavelengths in the Galaxy, LMC and SMC.}
\label{table:metal_pca}
\scalebox{1.0}{
\begin{tabular}{ccccc}
\hline
\hline
  & \multicolumn{2}{c}{A} & \multicolumn{2}{c}{B} \\
\hline
PC&	$\lambda$&	    C.P. &	$\lambda$&	  C.P.\\
\hline
      1&    52.7828&    52.7869&    45.6006&    45.6056\\
      2&    17.0784&    69.8666&    21.6212&    67.2292\\
      3&    11.5115&    81.3790&    10.5966&    77.8270\\
      4&     6.2954&    87.6749&     7.8642&    85.6920\\
      5&     4.0331&    91.7083&     4.1406&    89.8330\\
      6&     3.0631&    94.7716&     3.4460&    93.2794\\
      7&     1.3955&    96.1672&     1.7380&    95.0176\\
      8&     1.0315&    97.1987&     1.4948&    96.5126\\
      9&     0.8680&    98.0668&     1.1251&    97.6379\\
     10&     0.5596&    98.6265&     0.8645&    98.5025\\
\hline
\end{tabular}}
\end{center}
\end{table}

We also studied the Cepheid light curve structure using principal component analysis (PCA) method as an independent probe with respect to Fourier analysis. The PCA method is very useful to differentiate small features in the light curve structure of an ensemble analysis. The application of principal component analysis method to variable star light curves is discussed in detail in \citet{kanbur2002, kanbur2004, tanvir2005, deb09}. This method transforms the original correlation matrix of magnitudes obtained from light curves to a new set of elementary light curves and principal component coefficients. Let us consider a sample of n light curves such that $m^i$ represents the $i^{th}$ light curve with $p$ number of data points per light curve, then:

\begin{equation}
X^i_{j} = \frac{m^i_{j} - \overline{m^i}}{s^i\sqrt(n)},
\label{eq:pca1}
\end{equation}

\noindent is a $n\times p$ matrix of normalised magnitudes with $1<j\le p$. Also, $\overline{m^i}$ is the mean magnitude of the light curve and $s^i$ represents the standard deviation on the mean. Since the theoretical light curves are generated with multiple phase cycles, we   interpolate these light curves to obtain $p=100$ magnitudes between phase 0 and 1. The correlation matrix,

\begin{equation}
C_{jk} = \frac{1}{n} \sum_{i=1}^{n} X^i_{j} X^i_k,
\label{eq:pca2}
\end{equation}

\noindent measures the relationship between $j^{th}$ and $k^{th}$ data points averaged over all stars in the sample. We want to construct a Cepheid light curve as a linear combination of elementary light curves such that:

\begin{equation}
m(i) = \sum_{l=1}^{p} PC_{l}(i) E^l_i,
\label{eq:pca3}
\end{equation}

\noindent where, $PC_l(i)$ are the principal components and $E^l_i$ are the elementary light curves. These elementary light curves and their coefficients are determined using the input C matrix. The $E^l_i$ are orthogonal to each other and given by the solution of the eigenvalue equation, $CE=\lambda E$. The solution provides a set of eigenvectors $E^l_i$ and their corresponding eigenvalues $\lambda^l$. The percentage of total variation that can be accounted by $\lambda^l$ is given by $\lambda^l/\sum_{l} \lambda^l$. We can project each light curve on the eigenvectors such that, 

\begin{equation}
PC_l(i) = \sum_{j=1}^{p} X^i_{j} E^l_j.
\label{eq:pca4}
\end{equation}

If we consider all $p$ solutions of the eigenvalue problem then we can reproduce the light curve using equations~\ref{eq:pca3} and \ref{eq:pca4}. If we assume that first $q$ principal components are sufficient to reproduce the light curve of original $p$ data points, then we have a reduced ($n\times q$) matrix. Since, the majority of light curve information is contained in first few principal components, this method is very useful for reducing dimensionality in astronomical data analysis. We note that the Fourier decomposition method requires each star to be fitted individually for an optimum order of fit. Therefore, PCA method can be used to perform an ensemble analysis of large catalogues, provided the light curves are preprocessed to similar dimensions.

\begin{figure*}
\begin{center}
\includegraphics[width=1.0\textwidth,keepaspectratio]{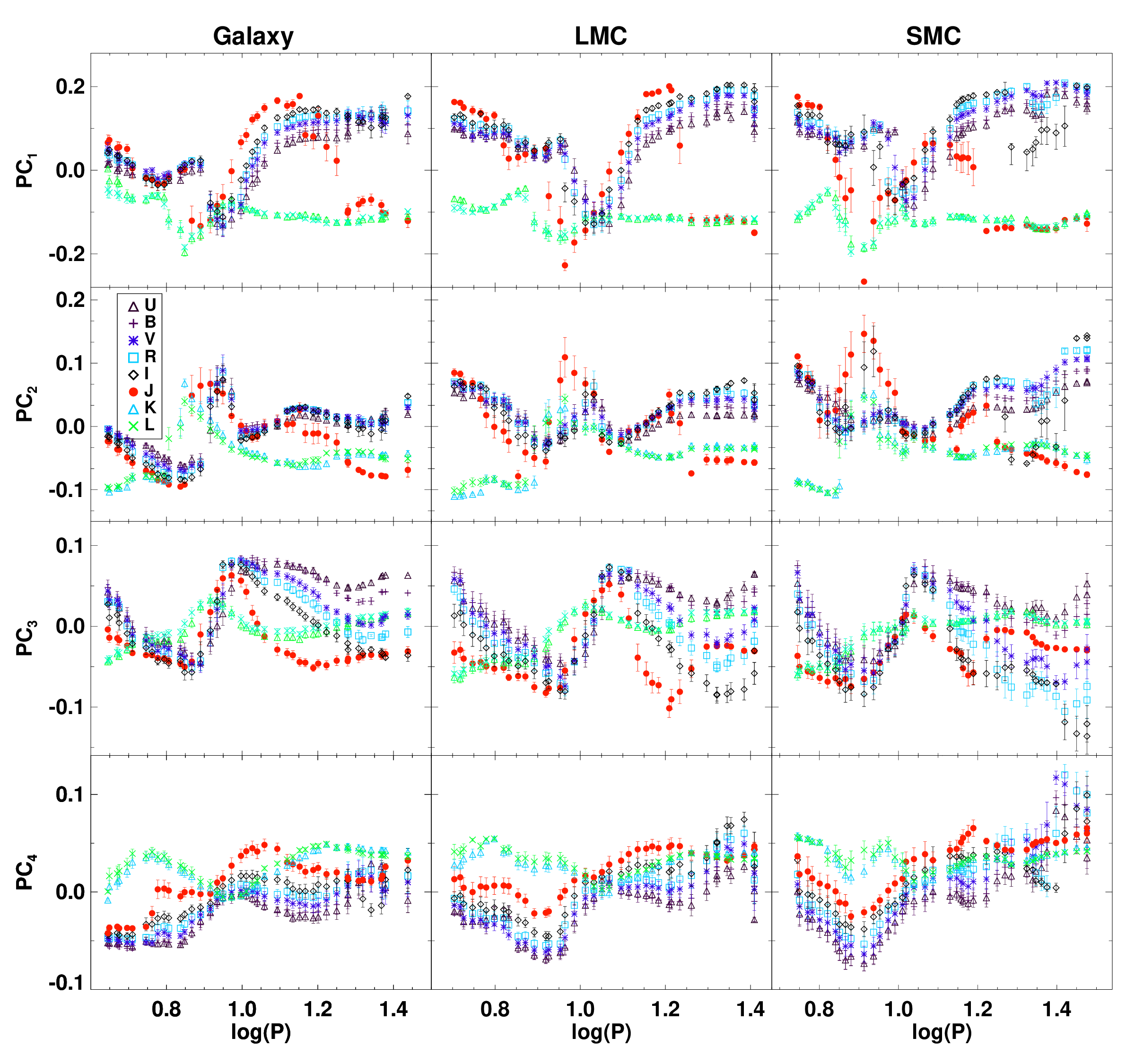}
\caption{Multi-wavelength mean principal components for the theoretical light curves of Cepheids with chemical compositions relative to the Galaxy, LMC and SMC. The error bars represent the standard deviation on the mean.}
\label{fig:mean_pca}
\end{center}
\end{figure*}

\begin{figure*}
\begin{center}
\includegraphics[width=1.0\textwidth,keepaspectratio]{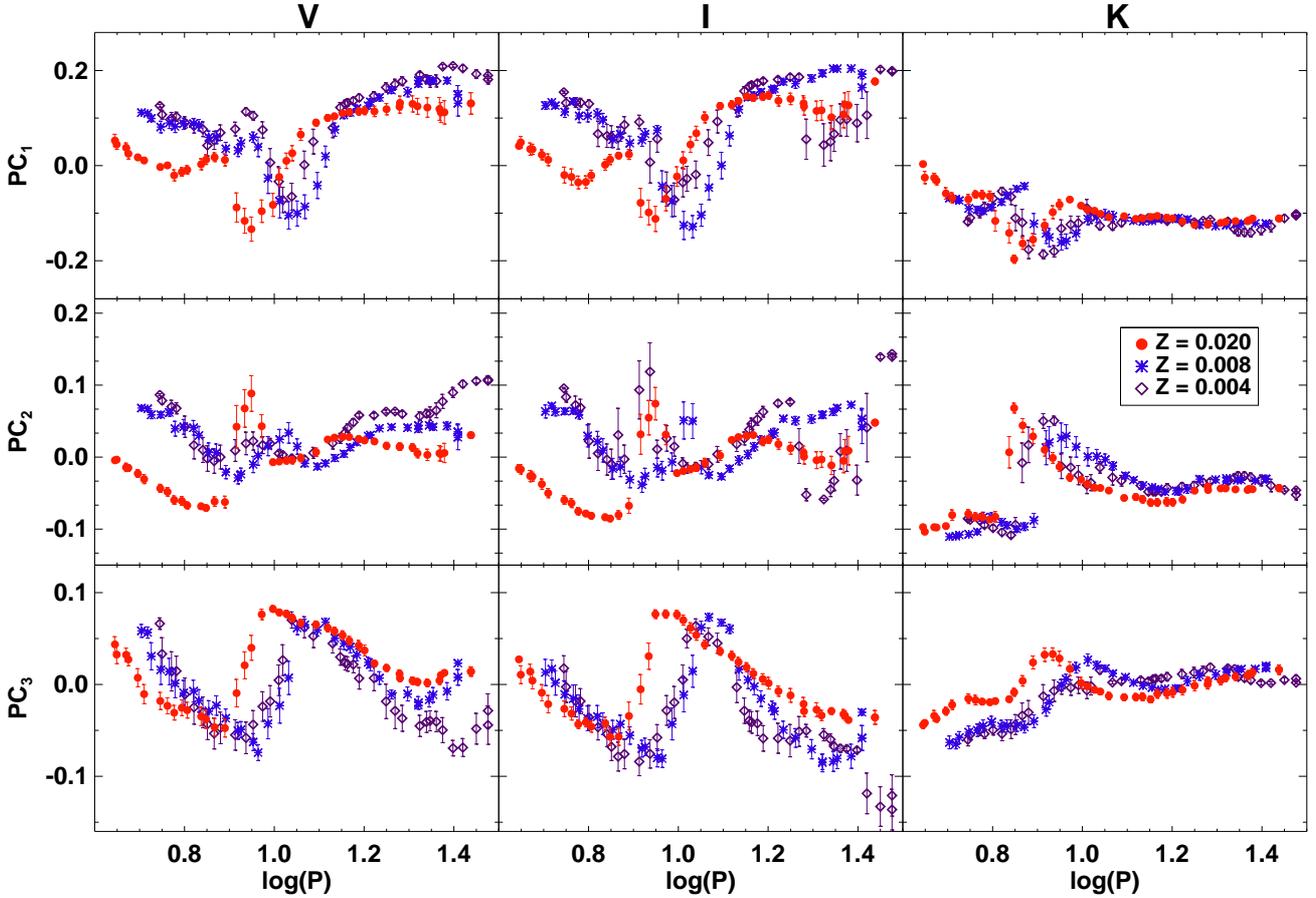}
\caption{A comparison of the mean principal components for FU Cepheids with different compositions.}
\label{fig:metal_pca}
\end{center}
\end{figure*}

\begin{figure*}
\begin{center}
\includegraphics[width=1.0\textwidth,keepaspectratio]{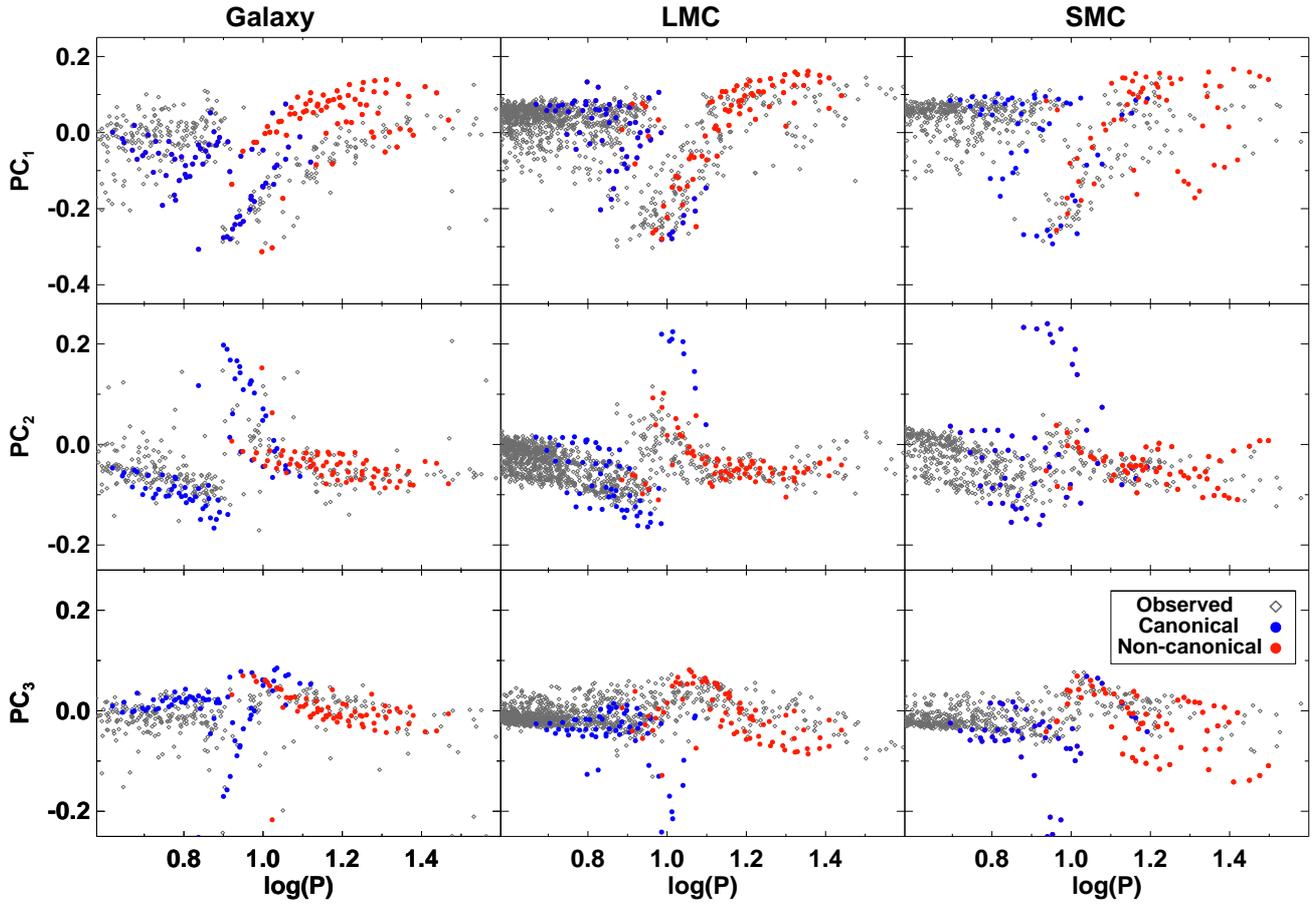}
\caption{A comparison of the theoretical and observed $I$-band principal components for FU Cepheids in the Galaxy and MC.}
\label{fig:compare_pca}
\end{center}
\end{figure*}

\begin{figure*}
\begin{center}
\includegraphics[width=1.0\textwidth,keepaspectratio]{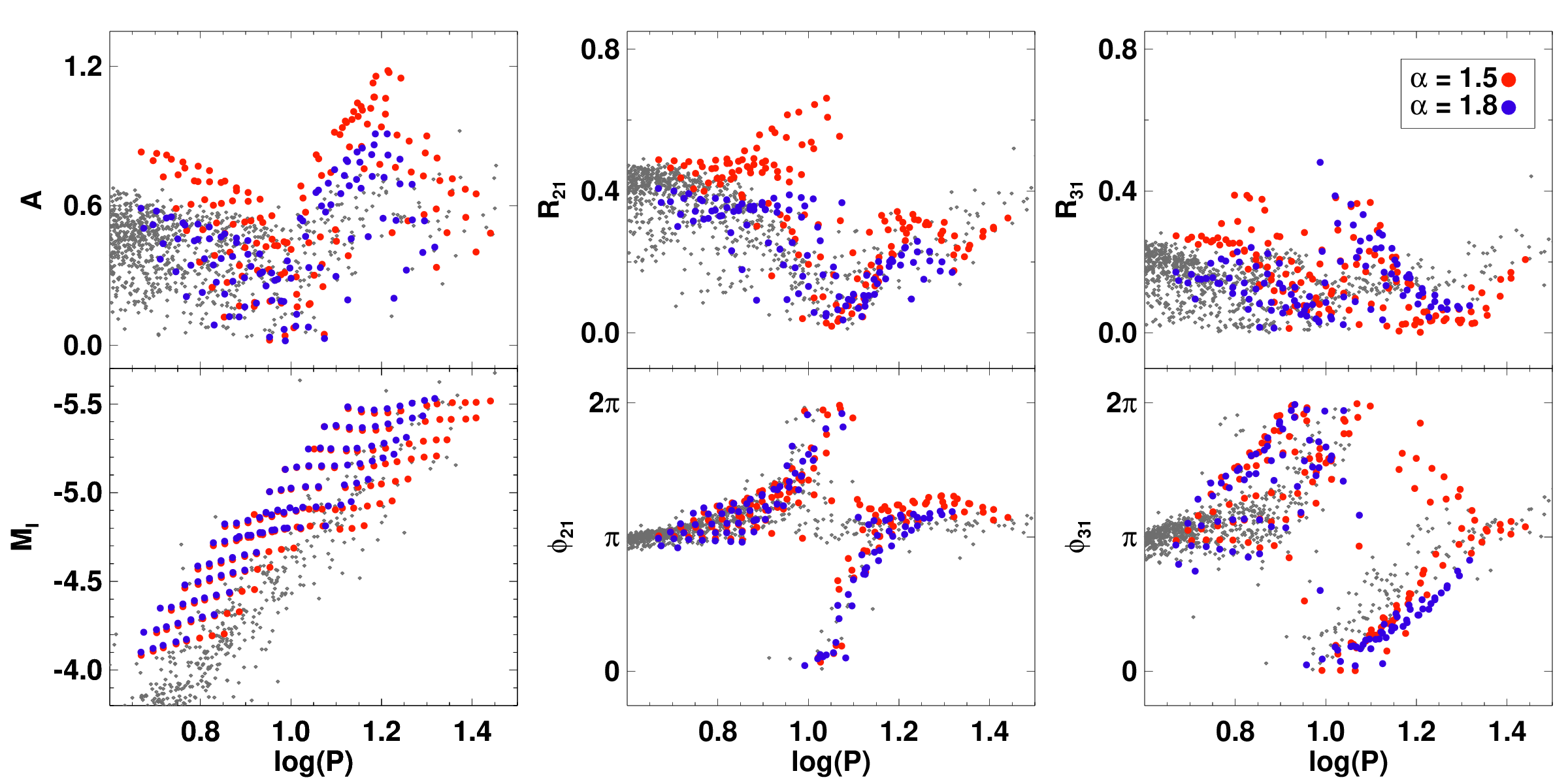}
\caption{A comparison of the theoretical light curve parameters with different mixing length parameter for metal abundance, Z=0.008. The grey dots represent observed parameters in $I$-band for OGLE LMC Cepheids.}
\label{fig:mixing_l}
\end{center}
\end{figure*}

\subsection{Multiwavelength principal components}

We carried out a PCA analysis on the ensemble of theoretical light curves for $BVRIJKL$ wavebands and for Cepheid models with Z=0.02, 0.008 and 0.004, representative of the Galaxy, LMC and SMC. Table~\ref{table:metal_pca} lists the first ten eigenvalues and their cumulative percentages for a PCA analysis to the multiband theoretical light curves of Cepheid variables. In order to observe the variation of principal components with wavelength we use sliding mean calculations as discussed in previous section. Fig.~\ref{fig:mean_pca} displays the variation of the mean of first four principal components as a function of period and wavelength for different chemical abundances. We find that first principal component shows two separate populations for optical and NIR wavelengths. The $KL$ bands are distinct from optical bands for the first principal component plots in all three galaxies. The $J$-band contributes to optical bands at shorter periods and shifts to $KL$ population for longest periods. The first principal component shows a clear resonance feature at 10 days \citep{antonello1994, kanbur2002, deb09} at optical wavelengths. The resonance feature in principal components varies at NIR wavelengths, reflecting an offset to earlier periods as compared to optical wavelengths. Similar resonance feature is also seen in second and third principal components. The first principal component also provide evidence of an increase with wavelength at optical bands for compositions relative to all three galaxies. Similar progression is also seen in the fourth principal component. However, no clear variation with wavelength is observed in case of the second and third principal components.

We also plot the mean of the first three principal components for compositions relative to the Galaxy, LMC and SMC in $VIK$-bands in Fig.~\ref{fig:metal_pca}. For short period ($P < 10$ days) Cepheids, the principal components increase with decreasing metallicity with a more pronounced progression for second principal component. This is similar to Fourier amplitude parameters. This progression is not observed at the long period ends. In case of $K$-band, the Galactic models with $P>10$~days occupy a different sequence in second principal component with respect to their lower metallicity counterparts in the MC. The third principal component show a decrease with metallicity for short period Cepheid models in $K$-band.

We also compare theoretical and observed principal components at optical wavelengths as we did for Fourier parameters with observed $I$-band data from OGLE IV. The PCA is performed on the observed and theoretical light curves simultaneously and the eigenvalues are listed in 
Table~\ref{table:metal_pca}. Fig.~\ref{fig:compare_pca} displays the first three principal components for the theoretical and observed 
$I$-band light curves for FU Cepheids in the Galaxy, LMC and SMC. We find a large offset between theory and observations in the first principal component for the Galactic Cepheid models, specifically for periods greater than 10 days. These are non-canonical set of models in the Galaxy, in contrast to the large offset in $R_{21}$ parameters for canonical models in the MC. Further, the short period canonical set of Galactic models also show a minimal offset for all principal components. The MC models are overall consistent with observations for first three principal components. Both the theoretical and observed principal components show similar resonance feature around 10 days. The third principal component shows a greater scatter at long period end for lower metallicity Cepheid models.

\section{Discussion and Conclusions}
\label{sec:discuss}

We studied the light curve parameters for the theoretical light curves of Cepheid variables as a function of period, wavelength and metallicity and compared those with observations. The motivation for performing a detailed quantitative comparison between observed and theoretical light curves is to provide strong constraints on stellar pulsation codes that incorporate stellar atmosphere models to produce Cepheid light curves in multiple bands.

We find that theoretical amplitudes are systematically larger than the observed amplitudes at optical wavelengths. One important variable that can affect amplitudes in theoretical models is the adopted mixing length parameter. It is well known that convection plays an important part in stellar pulsation, particularly at the red edge, where it acts to damp down growing perturbations by leaking energy out of the ionization zones around maximum compression. Increasing the mixing length parameter increases the average length over which gas pockets can transfer energy from hot to cool regions and, will act to increase the efficiency of convection and damp down pulsations. Thus, this parameter is very likely to lead to lower theoretical amplitudes. 

Fig.~\ref{fig:mixing_l} displays the result of increasing the mixing length parameter from 1.5 to 1.8 for a series of models with LMC composition covering the period range $0.6\le \log (P) \le 1.5$. We clearly see a decrease in the theoretical amplitudes sufficient to bring them into closer agreement with observations. There is very little change in the Fourier phase parameter values, especially for $\log (P) < 1$. In case of $\phi_{31}$, there were two branches for $\log (P) > 1.15$ with $\alpha = 1.5$. This unrealistic feature is no longer visible with $\alpha = 1.8$. Most importantly, the large discrepancy in $R_{21}$ parameter with respect to observations is solved with increased mixing length. This does not cause a significant change in amplitudes very close to 10 days but the $R_{21}$ values for models with $\log(P) > 1.2$ are also decreased. However, increasing this parameter also causes a zero-point offset in bolometric magnitudes, for example, the zero-point of the $I$-band period-luminosity relation is shifted. This may lead to a bias in distance estimates with theoretical P-L relations, depending on the choice of mixing length parameters \citep[see, ][ for a detailed discussion]{fiorentino2007}. These effects of varying various input parameters with wavelength and metallicity will be the subject of future studies.

Another important result of this paper is the potential of this method to discriminate between canonical and non-canonical models. A major
discrepancy between theory and observations occurs in the $R_{21}$ parameter for periods between $0.8\le \log (P) \le1.1$, for the LMC and SMC with $\alpha = 1.5$. These models are all canonical ML models. In contrast, the non-canonical models agree very well with observations in terms of this Fourier parameter. This is true for all three galaxies in both $V$ and $I$-bands. We have shown that this discrepancy can be solved with higher mixing length. Alternatively, we can also change a different parameter for the canonical models, for example, decreasing the mass and adopting brighter luminosity levels. \citet{bono2002} and \citet{fiorentino2007} have shown that increasing mixing length narrows the width of the instability strip as the red boundary gets bluer. So, there is change towards hotter temperatures which can be responsible to overcome the large offset in $R_{21}$, as we had shown that the discrepant Cepheid models have relatively lower temperatures. 

Some recent studies have used multidimensional hydrodynamical simulations to study the convection and a relation to mixing length theory \citep[for example,][]{Mundprecht2013, Mundprecht2015, magic2015, salaris2015}. Even if the adopted nonlinear models include a non local time-dependent treatment of convection, a free parameter related to the mixing length is adopted in the code in order to close the nonlinear system of dynamical and convective equations. According to some authors \citep[see,][]{Mundprecht2013, Mundprecht2015} a 2D approach is needed to perform a more realistic simulation of the complex coupling between pulsation and convection. On the basis of these simulations, even if our adopted convective flux formulation is in better agreement with the 2D results than other 1D formulations in the literature \citep[see,][for details]{Mundprecht2015}, some phase-dependent corrections might be needed in order to properly follow the evolution of the pulsation-convection coupling along the pulsation cycle.

A more detailed study can provide very stringent constraints discriminating between canonical and non-canonical ML relations used in stellar pulsation and inform how different parameters affect the changes in light curve structure. Given that these ML relations are based on stellar evolutionary calculations with differing input physics, this method has the potential to constrain both theories of stellar evolution and pulsation. \\

We summarize the results from this study as follows: 

\begin{itemize}

{\item We use full amplitude, nonlinear, convective hydrodynamical models to generate Cepheid light curves with chemical compositions representative of the Galaxy and Magellanic Clouds. We use Fourier decomposition and principal component analysis methods to analyse the  structure of observed and theoretical Cepheid light curves and discuss the variation with period, wavelength and metallicity.}

{\item We find that the theoretical amplitudes in optical bands are systematically larger than the observed amplitudes over the entire period range except in the vicinity of 10 days. The theoretical amplitudes can be decreased by increasing the mixing length parameter. This causes a zero-point offset in amplitude parameters, except close to the center of Hertzsprung progression, and in bolometric mean magnitudes.}

{\item Higher mass Galactic Cepheid models lead to an increase in amplitude parameters for period greater than 10 days. A decrease in phase parameters with increasing mass is observed for all wavelengths and metallicities. For a fixed composition, the Fourier amplitude parameters decrease with increasing wavelength while the phase parameters increase as a function of wavelength at a given period, similar to observed Fourier parameters.} 

{\item The $R_{21}$ parameter shows a large offset with respect to observations for short period ($0.8 \le \log (P) \le 1.1$) Cepheid models at optical wavelengths. This discrepancy occurs for canonical sets of models with masses greater than $M\geq6.0M_\odot$ and relatively lower temperatures in the range $5100 \leq T < 5400$. We note that the non-canonical mass-luminosity models display consistency with observations for this period range. Adopting a higher mixing length resolves this discrepancy as it shifts the red edge of the instability strip to hotter temperatures.}

{\item Theoretical mean $R_{21}$ parameter also exhibit a sharp rise around 20 days for wavelengths shorter than $J$ as compared to longer wavelengths, similar to amplitude parameters for observed light curves \citep{bhardwaj2015a}. The variation of mean Fourier parameters as a function of period and wavelength is consistent for both theoretical and observed light curves in most period bins, except for $R_{31}$.}

{\item We also observe two separate populations in the first principal component for optical and near-infrared bands with $J$-band contributing to both populations. The first principal components also show a clear resonance effect around 10 days at all wavelengths. The optical-band principal components for Magellanic Clouds Cepheids show a consistency between theory and observations but the first principal component displays a greater offset for Galactic Cepheid models.}

{\item The mean $R_{21}$, $R_{31}$ parameters and the first two principal components provide evidence of an increase with decreasing metallicity for $\log (P) < 1$. Further, the mean phase parameters also show the metallicity dependence with $\phi_{31}$ decreasing as a function of metallicity in most period bins. We also confirm that the central minimum of the Hertzsprung progression shifts to the longer periods with decrease/increase in metallicity/wavelength for both theoretical and observed light curves.}
\end{itemize}

Incorporating the quantitative information about light curve structure provided by Fourier fitting and PCA, we have detailed observational constraints on Cepheid light curve structure as a function of period, wavelength and composition. It is also possible to compare these with a smooth grid of models through modern machine learning methods \citep{bellinger2016} and obtain star by star estimates of $M$, $L$, $T_{eff}$, $X$ and $Z$, together with robust error estimates. This will be the subject of future work. We also demonstrate the need for more model calculations at short period ($\log (P) < 0.8$) end. The majority of observations with the greatest dispersion also lie shortward of this period cut and this also needs a further investigation.

\section*{Acknowledgments}
\label{sec:ackno}
AB acknowledges the Senior Research Fellowship grant 09/045(1296)/2013-EMR-I from the Human Resource Development Group (HRDG), which is a division of the Council of Scientific and Industrial Research (CSIR), India. We are grateful to G. Bono for encouraging and valuable discussions regarding stellar pulsation models.  HPS thanks University of Delhi for a R\&D grant. CCN thanks the funding from Ministry of Science and Technology (Taiwan) under the contract NSC104-2112-M-008-012-MY3. This work was initiated under the Indo-US Science and Technology Forum funded Indo-US joint networked centre on analysis of variable star data.

\bibliographystyle{mnras}
\bibliography{model_paper}

\appendix

\section{Asymmetry in Cepheid light curves}
\label{sec:asym}
\begin{figure*}
\centering
  \begin{tabular}{@{}cc@{}}
    \includegraphics[width=.42\textwidth]{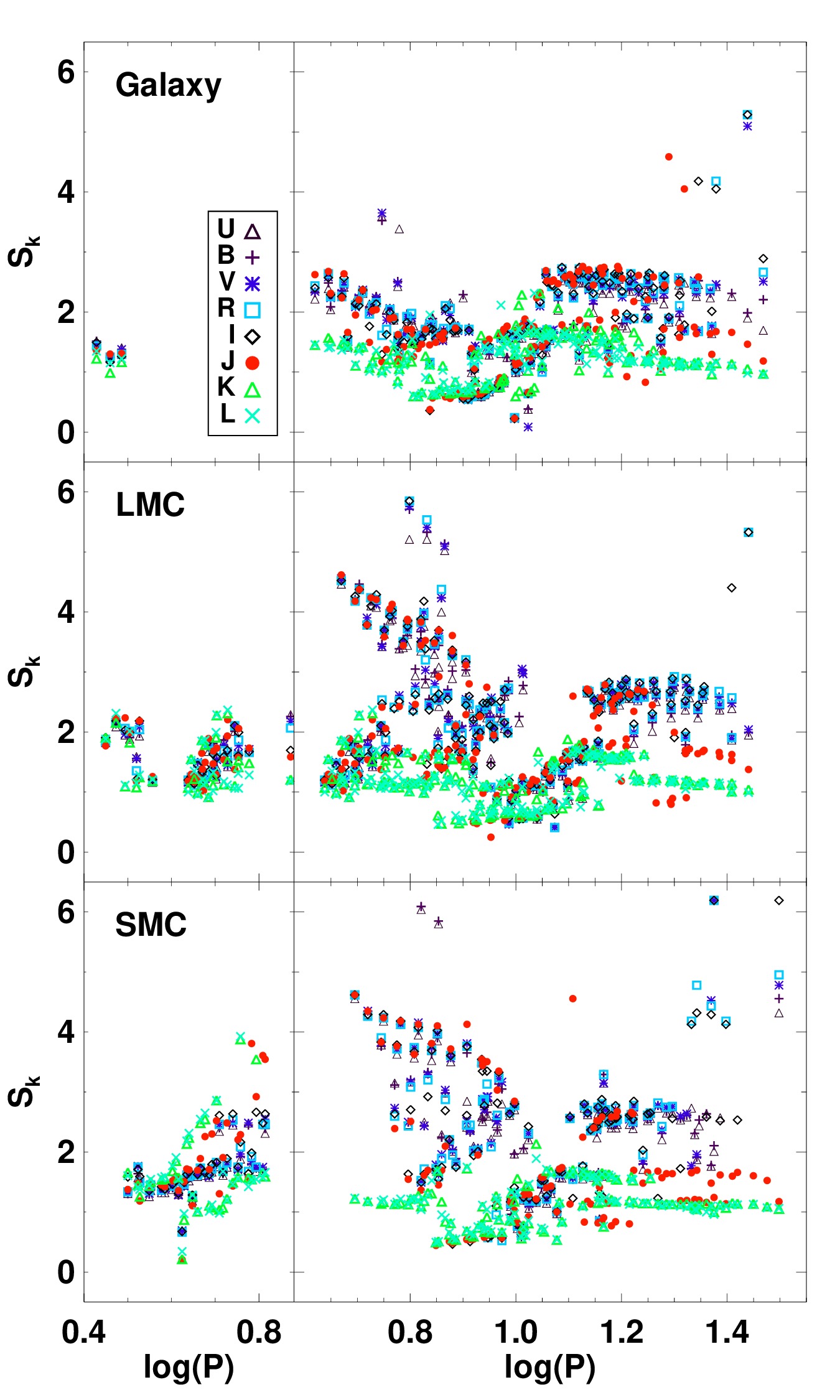} &
    \includegraphics[width=.42\textwidth]{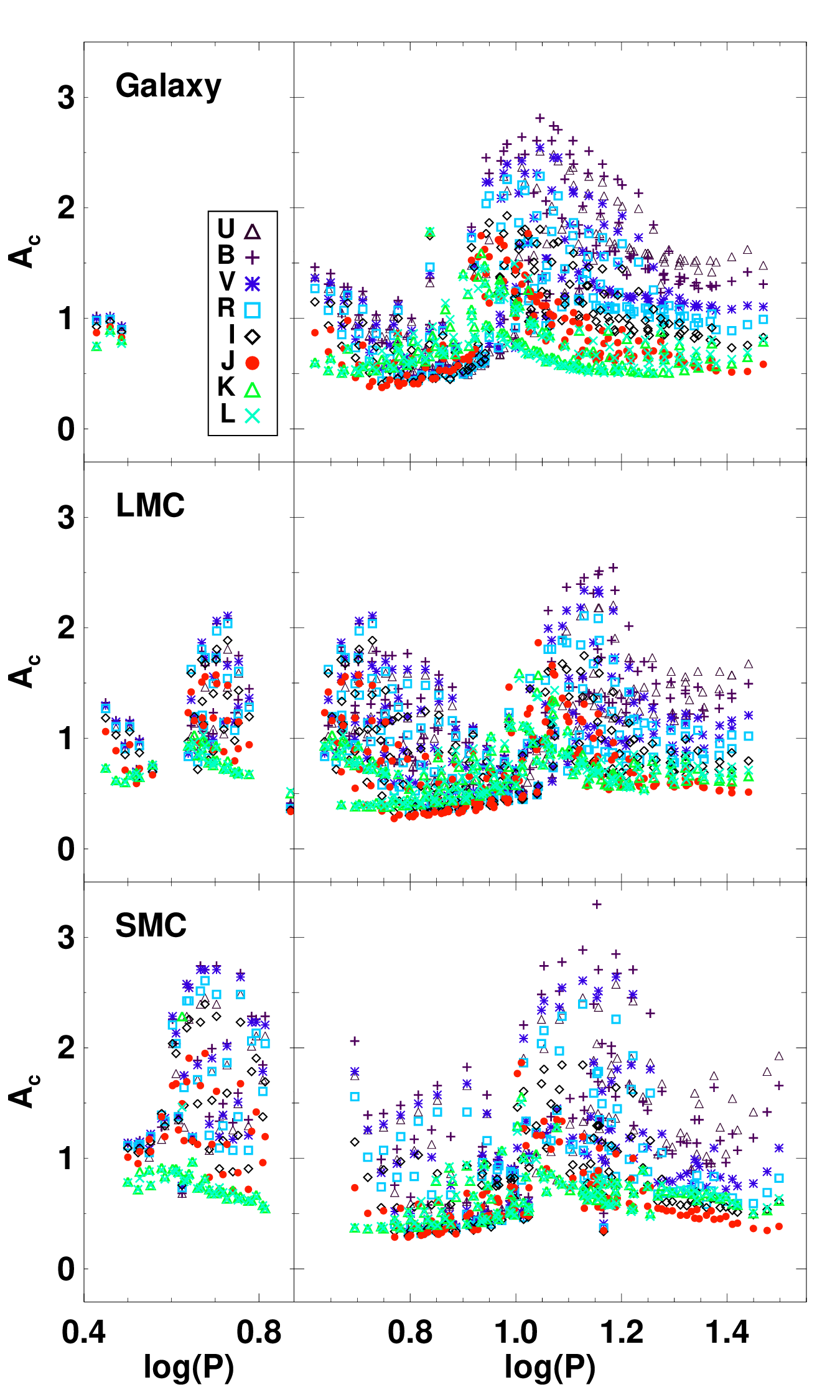}   \\
  \end{tabular}
 \caption{The variation of the skewness ($S_k$, left column) and acuteness ($A_c$, right column) parameters as a function of period and wavelength for the theoretical light curves of Cepheids with metal abundance, Z=0.02, Z=0.008 and Z=0.004. The right/left panels in each column display the FU/FO Cepheid models.}
 \label{fig:galaxy_skac}
\end{figure*}

We also discuss the asymmetry in the Cepheid light curves using skewness and acuteness parameters. The skewness is defined as the ratio of the phase durations of the descending branch to the rising branch. The acuteness quantifies the ratio of phase duration for the magnitudes fainter than median light to the brighter than median light \citep{bono2000d, bhardwaj2015a}. Fig.~\ref{fig:galaxy_skac} displays the variation of skewness and acuteness parameters as a function of period and wavelength for the theoretical light curves of FU and FO Cepheids with Z=0.02, Z=0.008 and Z=0.004. The optical and NIR populations are well separated in these skewness and acuteness planes. The skewness and acuteness parameters as a function of period also display different behaviour in optical with respect to NIR wavelengths. They attain a value close to unity at longer wavelengths suggesting the light curves are more symmetric, sinusoidal and flat-topped. 

At optical wavelengths, FU Cepheids with shorter periods have larger skewness in MC Cepheid models than the Galaxy, suggesting that theoretical light curves exhibit more left/right asymmetry for lower metallicities. The larger acuteness value for longer period Cepheids suggests that their light curves exhibit top/bottom asymmetry, for all compositions. Acuteness decreases with increase in wavelength for long period Galactic Cepheid models and this trend becomes less pronounced for lower metallicity models. FO Cepheid models show a smaller skewness and a larger acuteness values for MC Cepheid models. We recall that the observed skewness and acuteness parameters also show a decrease with increasing wavelength and Cepheids with 10 days periods attain a value close to unity \citep{bhardwaj2015a}.

\section{Fourier Interrelations}

\begin{figure}
\begin{center}
\includegraphics[width=0.5\textwidth,keepaspectratio]{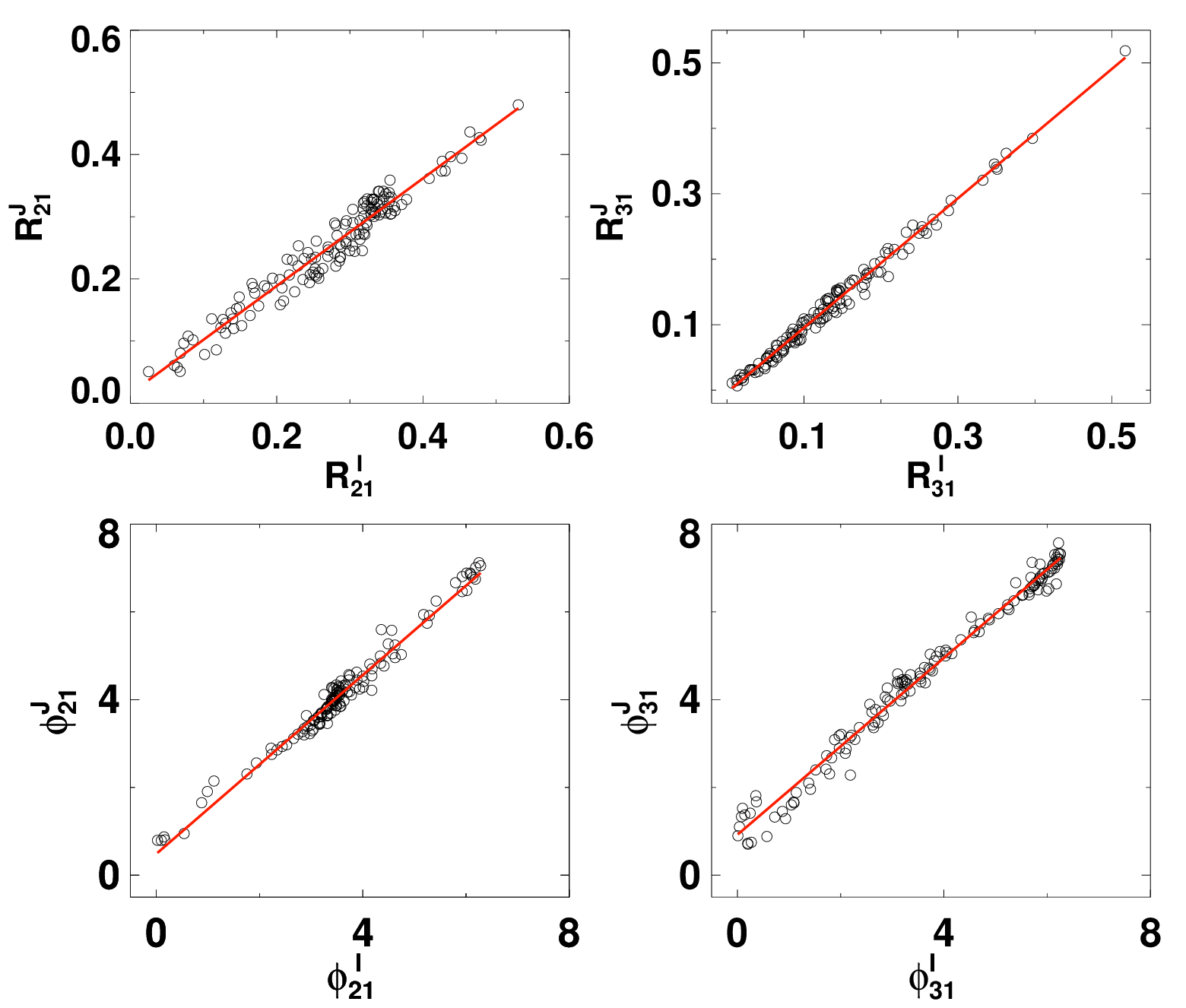}
\caption{The Fourier interrelations from $I$-$J$ bands with chemical compositions, Z=0.02.}
\label{fig:inter_relations}
\end{center}
\end{figure}

The high quality light curve data in the near-infrared is not available for Cepheids in the literature to determine very precise Fourier parameters. Although, there are some discrepancies between theoretical and observed light curve parameters, specifically at the short period end, there is an overall consistency in the variation of theoretical Fourier parameters with period and wavelength. Therefore, we derive the Fourier interrelations \citep[for example, in][]{ngeow2003} to obtain the Fourier parameters in various bands. The following equations :
\vspace{-2pt}

\begin{gather}
R_{21}^{\lambda_2} = \alpha + \beta R_{21}^{\lambda_1},~
R_{31}^{\lambda_2} = \alpha + \beta R_{31}^{\lambda_1},~
R_{41}^{\lambda_2} = \alpha + \beta R_{41}^{\lambda_1},\nonumber \\
\phi_{21}^{\lambda_2} = \alpha + \beta \phi_{21}^{\lambda_1},~
\phi_{31}^{\lambda_2} = \alpha + \beta \phi_{31}^{\lambda_1},~
\phi_{41}^{\lambda_2} = \alpha + \beta \phi_{41}^{\lambda_1},\nonumber 
\end{gather}

\noindent where $\lambda_1 < \lambda_2$, provide transformations for Fourier parameters between different wavelength bands. We derive separate set of equations for the chemical compositions relative to the Galaxy, LMC and SMC. The results are listed in 
Table~\ref{table:inter_relations}. A representative plot of the Fourier interrelations from $I$-$J$ bands with chemical compositions relative to the Galaxy is shown in Fig.~\ref{fig:inter_relations}. The optical interrelations show a strong correlations between Fourier parameters while there is more scatter in near-infrared bands. The internal dispersion increases for higher order phase parameters but we note that most characteristic features of the light curves are stored in the lower order Fourier parameters.

\begin{table*}
\begin{center}
\caption{The Fourier interrelations in multiple bands for Galactic, LMC and SMC models.}
\label{table:inter_relations}
\scalebox{0.97}{
\begin{tabular}{ccccccccccc}
\hline
\hline
$\lambda_1-\lambda_2$& 	FP&	 \multicolumn{3}{c}{Z=0.02}&\multicolumn{3}{c}{Z=0.008}&	\multicolumn{3}{c}{Z=0.004}\\

		&	&	$\beta$ & $\alpha$ & $\sigma$&	$\beta$ & $\alpha$ & $\sigma$&	$\beta$ & $\alpha$ & $\sigma$  \\
\hline
\hline
         V$-$I&       $R_{21}$&      0.989$\pm$0.009     &     -0.004$\pm$0.005     &      0.011&      0.997$\pm$0.006     &     -0.012$\pm$0.004     &      0.011&      0.975$\pm$0.011     &     -0.014$\pm$0.007     &      0.020\\
              &       $R_{31}$&      1.024$\pm$0.006     &     -0.004$\pm$0.002     &      0.007&      1.000$\pm$0.006     &     -0.003$\pm$0.002     &      0.006&      0.973$\pm$0.010     &     -0.005$\pm$0.004     &      0.009\\
              &       $R_{41}$&      0.979$\pm$0.005     &      0.002$\pm$0.002     &      0.003&      0.967$\pm$0.003     &      0.003$\pm$0.001     &      0.002&      0.915$\pm$0.006     &      0.006$\pm$0.002     &      0.004\\
              &    $\phi_{21}$&      1.008$\pm$0.003     &      0.255$\pm$0.006     &      0.037&      1.000$\pm$0.007     &      0.336$\pm$0.015     &      0.080&      1.036$\pm$0.015     &      0.289$\pm$0.030     &      0.122\\
              &    $\phi_{31}$&      1.002$\pm$0.003     &      0.478$\pm$0.007     &      0.059&      0.987$\pm$0.005     &      0.596$\pm$0.012     &      0.105&      0.992$\pm$0.008     &      0.692$\pm$0.022     &      0.175\\
              &    $\phi_{41}$&      1.002$\pm$0.003     &      0.670$\pm$0.008     &      0.071&      0.993$\pm$0.005     &      0.784$\pm$0.013     &      0.119&      1.007$\pm$0.012     &      0.923$\pm$0.029     &      0.214\\

         I$-$J&       $R_{21}$&      0.865$\pm$0.018     &      0.016$\pm$0.010     &      0.021&      0.909$\pm$0.013     &     -0.001$\pm$0.008     &      0.024&      0.854$\pm$0.021     &      0.002$\pm$0.013     &      0.037\\
              &       $R_{31}$&      0.989$\pm$0.009     &     -0.004$\pm$0.003     &      0.010&      0.921$\pm$0.008     &      0.000$\pm$0.003     &      0.009&      0.888$\pm$0.015     &     -0.002$\pm$0.007     &      0.015\\
              &       $R_{41}$&      0.997$\pm$0.006     &     -0.002$\pm$0.002     &      0.004&      0.898$\pm$0.006     &      0.004$\pm$0.002     &      0.004&      0.812$\pm$0.014     &      0.009$\pm$0.006     &      0.009\\
              &    $\phi_{21}$&      1.019$\pm$0.013     &      0.485$\pm$0.030     &      0.191&      1.016$\pm$0.017     &      0.415$\pm$0.036     &      0.194&      0.999$\pm$0.021     &      0.496$\pm$0.045     &      0.194\\
              &    $\phi_{31}$&      1.008$\pm$0.010     &      0.921$\pm$0.028     &      0.236&      1.007$\pm$0.012     &      0.822$\pm$0.030     &      0.269&      1.020$\pm$0.015     &      0.857$\pm$0.037     &      0.301\\
              &    $\phi_{41}$&      0.980$\pm$0.013     &      1.388$\pm$0.034     &      0.314&      0.972$\pm$0.015     &      1.259$\pm$0.039     &      0.339&      0.974$\pm$0.022     &      1.351$\pm$0.054     &      0.397\\

         J$-$K&       $R_{21}$&      0.588$\pm$0.030     &      0.023$\pm$0.015     &      0.031&      0.596$\pm$0.031     &      0.016$\pm$0.017     &      0.052&      0.570$\pm$0.041     &      0.021$\pm$0.023     &      0.063\\
              &       $R_{31}$&      0.595$\pm$0.010     &      0.008$\pm$0.004     &      0.010&      0.608$\pm$0.017     &      0.003$\pm$0.007     &      0.018&      0.556$\pm$0.029     &      0.003$\pm$0.012     &      0.026\\
              &       $R_{41}$&      0.629$\pm$0.008     &      0.002$\pm$0.003     &      0.005&      0.544$\pm$0.014     &      0.008$\pm$0.005     &      0.009&      0.426$\pm$0.025     &      0.015$\pm$0.009     &      0.014\\
              &    $\phi_{21}$&      0.859$\pm$0.011     &      1.245$\pm$0.025     &      0.158&      0.897$\pm$0.024     &      1.093$\pm$0.054     &      0.293&      0.763$\pm$0.031     &      1.515$\pm$0.070     &      0.307\\
              &    $\phi_{31}$&      1.071$\pm$0.011     &      1.396$\pm$0.027     &      0.241&      1.069$\pm$0.017     &      1.345$\pm$0.041     &      0.332&      1.061$\pm$0.018     &      1.148$\pm$0.043     &      0.327\\
              &    $\phi_{41}$&      1.040$\pm$0.016     &      2.189$\pm$0.036     &      0.283&      0.992$\pm$0.015     &      2.201$\pm$0.036     &      0.285&      0.956$\pm$0.016     &      2.111$\pm$0.040     &      0.282\\

\hline
\end{tabular}}
\end{center}
\end{table*}

\section{Multiband Theoretical Fourier parameters}

\begin{figure*}
\begin{center}
\includegraphics[width=0.9\textwidth,keepaspectratio]{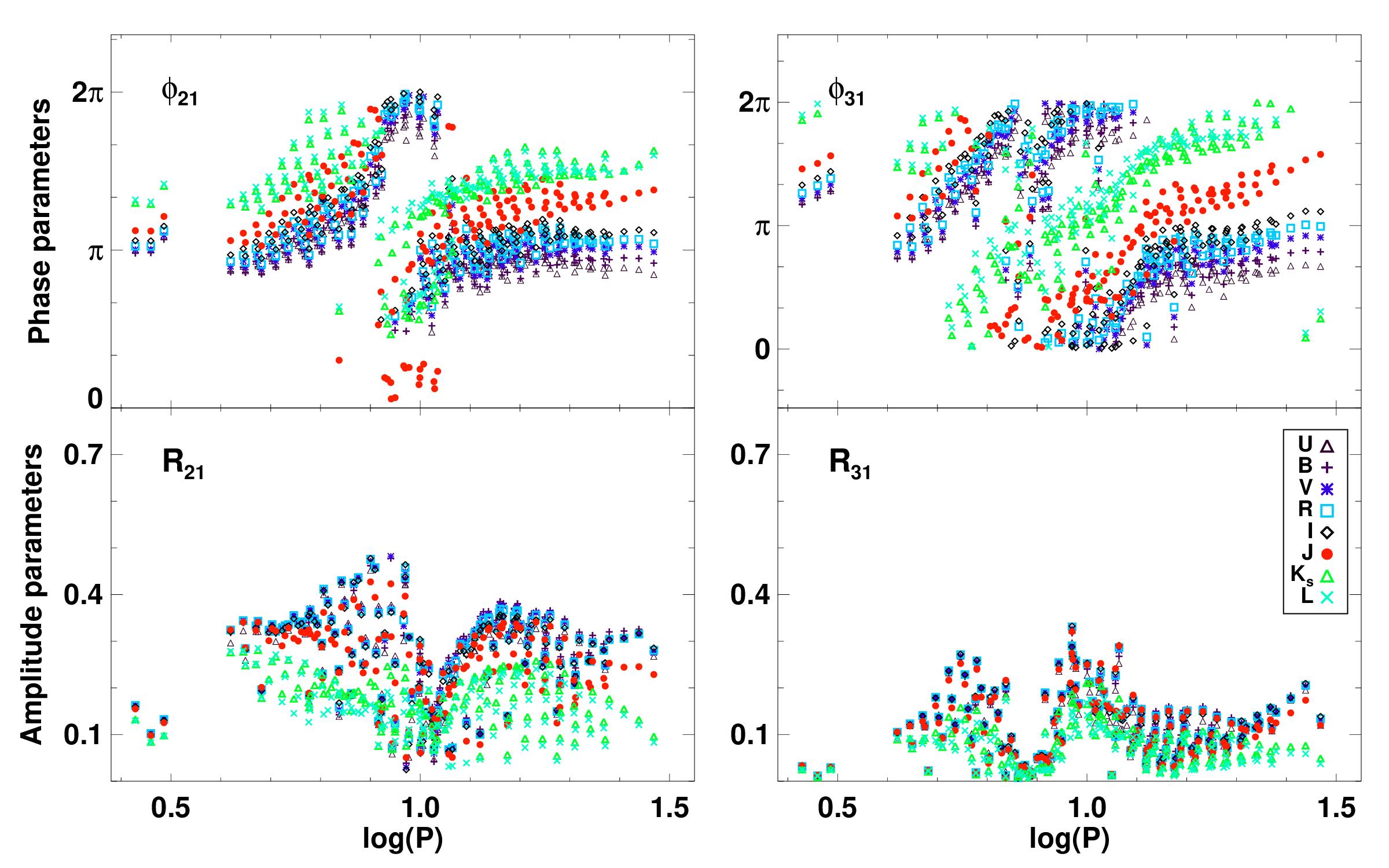}
\caption{Multi-wavelength Fourier parameters for the theoretical light curves of Cepheids with metal abundance, Z=0.02.
Some of the $\phi_{31}$ parameters are offset by $2\pi$ for plotting purposes. The models with $\log (P) < 0.5$ represent the FO Cepheids. We applied 2.5$\sigma$ clipping of a $5^{th}$ order polynomial-fit in individual bands before plotting, for the ease of visual inspection of different progressions. This procedure removes only a few extreme outlier Cepheid models on the Fourier plane.}
\label{fig:z02y28_multiband}
\end{center}
\end{figure*}

\begin{figure*}
\centering
  \begin{tabular}{@{}cc@{}}
    \includegraphics[width=0.95\textwidth]{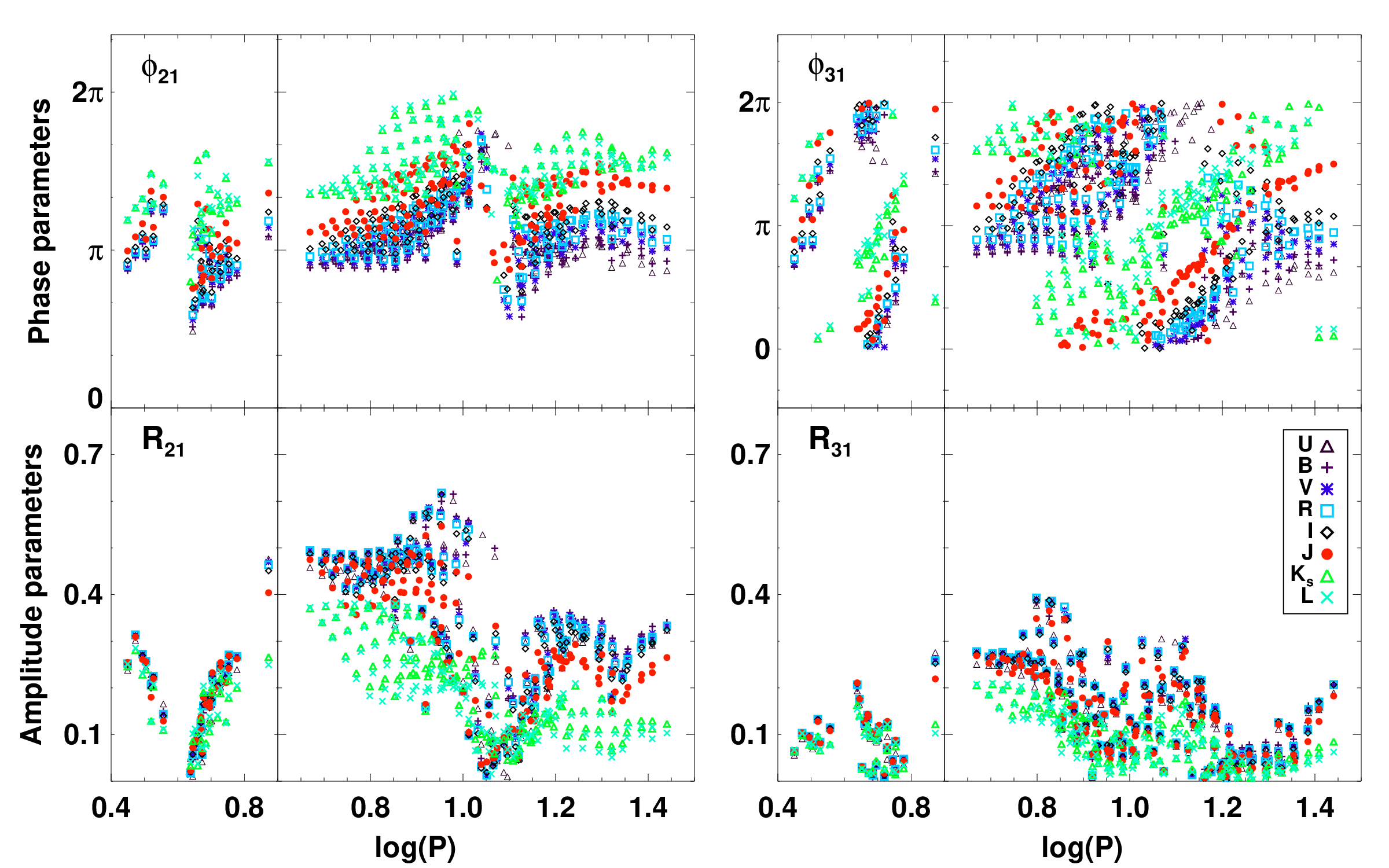} \\
    \includegraphics[width=0.95\textwidth]{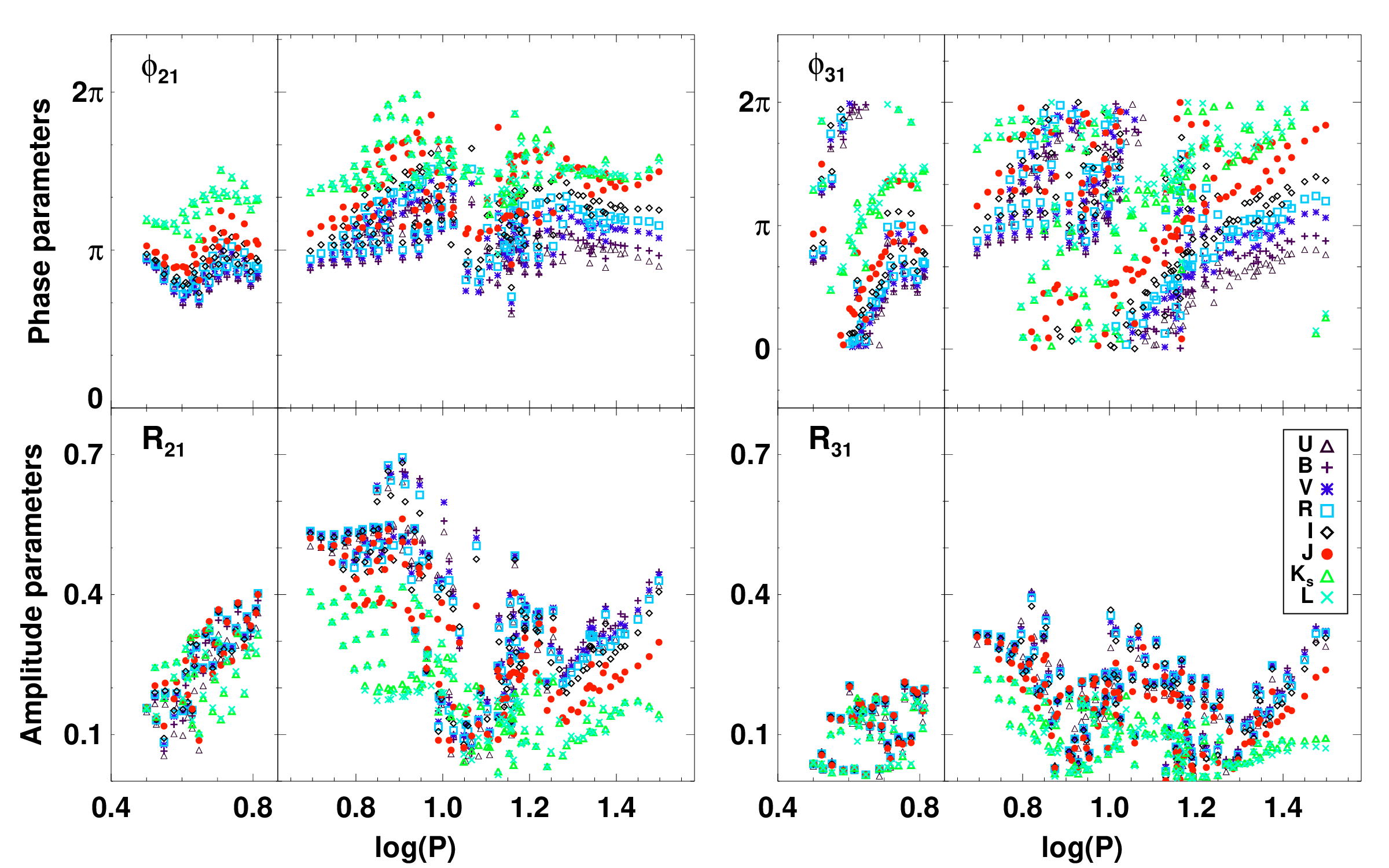}   \\
  \end{tabular}
 \caption{Same as Fig.~\ref{fig:z02y28_multiband} but for metal abundance, Z=0.008 (top) and Z=0.004 (bottom). The right/left panel in each column displays the FU/FO Cepheid models.}
 \label{fig:mc_multiband}
\end{figure*}

Figs.~\ref{fig:z02y28_multiband} and ~\ref{fig:mc_multiband} display the multiwavelength Fourier parameters for theoretical light curves of Cepheids in the Galaxy and Magellanic Clouds, respectively. 

\end{document}